\documentclass[%
unsortedaddress,
twocolumn,
 amsmath,amssymb,
 aps,
prb,
]{revtex4-1}

\usepackage{graphicx}
\usepackage{dcolumn}
\usepackage{bm}
\usepackage{multirow}
\usepackage[normalem]{ulem}

\usepackage{xcolor}

\usepackage{gensymb}

\graphicspath{{./figures/}}

\begin{document}

\title{Assessing the accuracy of compound formation energies with quantum Monte Carlo}
\author{Eric B. Isaacs}
\affiliation{Department of Materials Science and Engineering, Northwestern University, Evanston, Illinois 60208, USA}
\affiliation{Northwestern Argonne Institute of Science and Engineering, Evanston, Illinois 60208, USA}
\altaffiliation{Present address: HRL Laboratories LLC, Malibu, California 90265, USA}
\author{Hyeondeok Shin}
\affiliation{Computational Science Division, Argonne National Laboratory, Argonne, Illinois 60439, USA}
\author{Abdulgani Annaberdiyev}
\affiliation{Department of Physics, North Carolina State University, Raleigh, North Carolina 27695-8202, USA}
\affiliation{Center for Nanophase Materials Sciences Division, Oak Ridge National Laboratory, Oak Ridge, Tennessee 37831, USA}
\author{Chris Wolverton}
\affiliation{Department of Materials Science and Engineering, Northwestern University, Evanston, Illinois 60208, USA}
\author{Lubos Mitas}
\affiliation{Department of Physics, North Carolina State University, Raleigh, North Carolina 27695-8202, USA}
\author{Anouar Benali}
\affiliation{Computational Science Division, Argonne National Laboratory, Argonne, Illinois 60439, USA}
\author{Olle Heinonen}
\email{olle.heinonen@seagate.com}
\affiliation{Material Science Division, Argonne National Laboratory, Argonne, Illinois 60439, USA}
\affiliation{Northwestern Argonne Institute of Science and Engineering, Evanston, Illinois 60208, USA}
\altaffiliation{Present and permanent address: Seagate Technology, 7801 Computer Ave, Bloomington, MN 55435}

\date{\today}

\begin{abstract}
Accurately predicting the formation energy of a compound, which describes its thermodynamic stability, is a key challenge in materials physics. Here, we employ many-body quantum Monte Carlo (QMC) with single-reference trial functions to compute the formation energy of two electronically disparate compounds, the intermetallic VPt$_2$ and the semiconductor CuI, for which standard density functional theory (DFT) predictions using both the Perdew-Burke-Ernzerhof (PBE) and the strongly constrained and appropriately normed (SCAN) density functional approximations deviate markedly from available experimental values.
For VPt$_2$, we find an agreement between QMC, SCAN, and PBE0 estimates, which therefore remain in disagreement
with the much less exothermic experimental value.
For CuI, the QMC result agrees with neither SCAN nor PBE pointing towards DFT exchange-correlation  biases, likely related to the localized Cu $3d$ electrons.
Compared to the behavior of some density functional approximations within DFT, spin-averaged QMC exhibits a smaller but still appreciable deviation when compared to experiment. The QMC result is slightly improved by  incorporating spin-orbit corrections for CuI and solid I$_2$, so that experiment and theory are brought into imperfect but reasonable agreement within about 120~meV/atom.
\end{abstract}


\maketitle

\section{Introduction}\label{sec:intro}

Achieving reliable predictions of the thermodynamic properties of
solids is a critical challenge in materials physics. In particular, the
formation enthalpy of a compound with respect to its constituent
elements, $\Delta H_f$, is a key quantity that encapsulates phase
stability at zero temperature and strongly impacts the
finite-temperature phase diagram. Given that the thermodynamic stability of
inorganic crystalline compounds depends on small differences in  
energies of only 60~meV/atom on
average,\cite{bartel_critical_2020} highly accurate $\Delta H_f$
predictions are desirable. They may even be essential for cases
without the prospect of strong error cancellation, e.g., when a
compound's competing phases include elements and/or compounds with
very different electronic properties.

The challenge in predicting formation enthalpies ultimately derives
from our lack of a highly accurate and computationally efficient
electronic structure method, especially for solids. In the standard
approach of Kohn-Sham density functional theory
(DFT),\cite{hohenberg_inhomogeneous_1964,kohn_self-consistent_1965}
the challenging many-electron nature of the problem is circumvented by
(1) building the theory around a ground-state energy and electron
density derived from auxiliary non-interacting electrons and (2)
making non-systematically-improvable approximations to describe the
exchange and correlation interactions. Despite its many successes and
widespread usage,\cite{jones_density_2015,marzari2021electronic} the $\Delta H_f$ errors for
solids using DFT with common exchange-correlation approximations provide mixed results.
\cite{stevanovic_correcting_2012,kirklin_open_2015,zhang_efficient_2018,bartel_role_2019,friedrich_coordination_2019}
In particular, mean absolute $\Delta H_f$ errors of 259 and 110
meV/atom were found previously for the widely-used PBE
(Perdew-Burke-Ernzerhof) generalized-gradient approximation
(GGA)\cite{perdew_generalized_1996} and the newer and more
sophisticated SCAN (strongly constrained and appropriately normed)
meta-GGA,\cite{sun_strongly_2015}
respectively.\cite{isaacs_performance_2018}

In contrast to DFT, quantum Monte Carlo (QMC) is based on explicit many-electron wave functions, and it treats the exchange and correlation
effects directly by solving the stationary Schr\"odinger equation \cite{austin_quantum_2012}. In order
to cope with the vastly increased computational cost, QMC employs
stochastic algorithms, introducing statistical error bars, and often
requires substantial supercomputer resources. The use of QMC for
solids, as opposed to finite systems, is particularly computationally
demanding given the need to eliminate finite-size errors via (1)
employing supercells, and (2) considering multiple wavefunction phases,
or ``twists.''\cite{foulkes_quantum_2001,kolorenc_applications_2011}
There are further possible biases involved that can be divided into technical ones and more fundamental ones. The technical biases involve basis sets and 
wave function projection parameters, and in this study we deem these issues as being under control or marginal (see Appendices for more information). We also want to mention that we employ pseudopotentials which involve technical aspects, such as valence space fidelity to all-electron ion(s) and transferability, but also deeper issues due to non-local character of corresponding operators as opposed to the local nature of the original Hamiltonian. We have focused here only on the accuracy aspect and we have found the employed pseudopotentials to be acceptable for providing the energy differences that we are interested in with sufficient accuracy. On the fundamental side is the restriction to employ single-reference trial functions which affect the well-known fixed-node/phase bias
\cite{toulouse_chapter_2016,foulkes_quantum_2001,kolorenc_applications_2011,melton_spin-orbit_2016}. This aspect is significantly more challenging to control although systematic analysis and cross-comparisons between systems have proved useful in establishing quantitative bounds even for these errors \cite{annaberdiyev_cohesion_2021}. In addition, there is a significant effort  underway to establish a more routine use of multi-reference trial functions for solids in future \cite{benali_toward_2020}.

In this work, we employ QMC methods to compute $\Delta H_f$ and compare the results with DFT
and experimental measurements for crystals that represent two classes of materials. 
The first system of interest is VPt$_2$, which belongs to intermetallics with weaker bonding and with a significant spin-orbit effect originating from the Pt atom.  
It is well-known that metals pose additional challenges to QMC methods due to the presence of the Fermi surface and possible impact of long-range electron-electron correlations\cite{reboredo_optimized_2005}.
The second compound of interest is insulating CuI that involves spin-orbit effects from the iodine atoms. Since such systems have not been studied by QMC previously, our key goal has been to understand the {\em  feasibility and accuracy } of such calculations. In particular, we wanted to probe for the use of single-reference trial wave functions 
that represents the ``standard model" for QMC calculations.
Furthermore, for both of these systems,  
DFT-predicted values for $\Delta H_f$ deviate from the experimental data quite significantly, so that
another goal was to gain insights
into the DFT discrepancies.

Recently, the applicability of QMC has been expanding to systems with spin-orbit (SO) interactions based on developments of the fixed-phase spin-orbital diffusion Monte Carlo (FPSODMC) method that employs many-body wave functions built with two-component spinors\cite{melton_quantum_2016, melton_spin-orbit_2016}.
In particular, atomic and molecular systems have been studied in detail, revealing the significant impact of explicit spin-orbit effects on ordering of atomic and molecular excitations and other quantities such as binding energies \cite{melton_quantum_2016, melton_spin-orbit_2016, melton_projector_2019}. 
Therefore, the spin-orbit effect on formation energies is another point of interest of this work.

For the intermetallic VPt$_2$,
we find that QMC's prediction is closer to SCAN's than to PBE's, which clearly shows the difference 
between these two density functional approximations (DFAs) for a metal with weak bonds.
However, the difference between the theory and experiment still remains and its root cause is unclear. Experiments
on these systems are notoriously difficult, while on the theory side, the single-reference QMC 
could be reaching its accuracy limit here.
Nevertheless, this discrepancy, which is now exhibited by two independent theory approaches,
should provide a motivation for revisiting this system also on the experimental side.

For our second material, CuI,  the DFT formations energies using SCAN and PBE DFAs become nearly identical and they are significantly smaller in magnitude than the experimental value. Compared to DFT, QMC with averaged spin-orbit exhibits a much more exothermic formation energy prediction with reduced but still non-negligible error compared to experiment. We find that including the spin-orbit corrections for solid I$_2$ as well as CuI provides further mild improvement. After this correction, the QMC formation energy is about 120 ~meV/atom larger than the experiment, which is significantly better than the 
DFT SCAN and PBE values that underestimate the formation energy by some 
210~meV/atom with the most probable reason being poor description of Cu $3d$ states. Indeed, our probe for this effect with the hybrid PBE0 DFA indicates correction in the right direction and confirms the overall size of the spin-orbit effect as well.

\section{Description of systems}\label{sec:systems}

\begin{figure}[htbp]
  \centering
  \includegraphics[width=\linewidth]{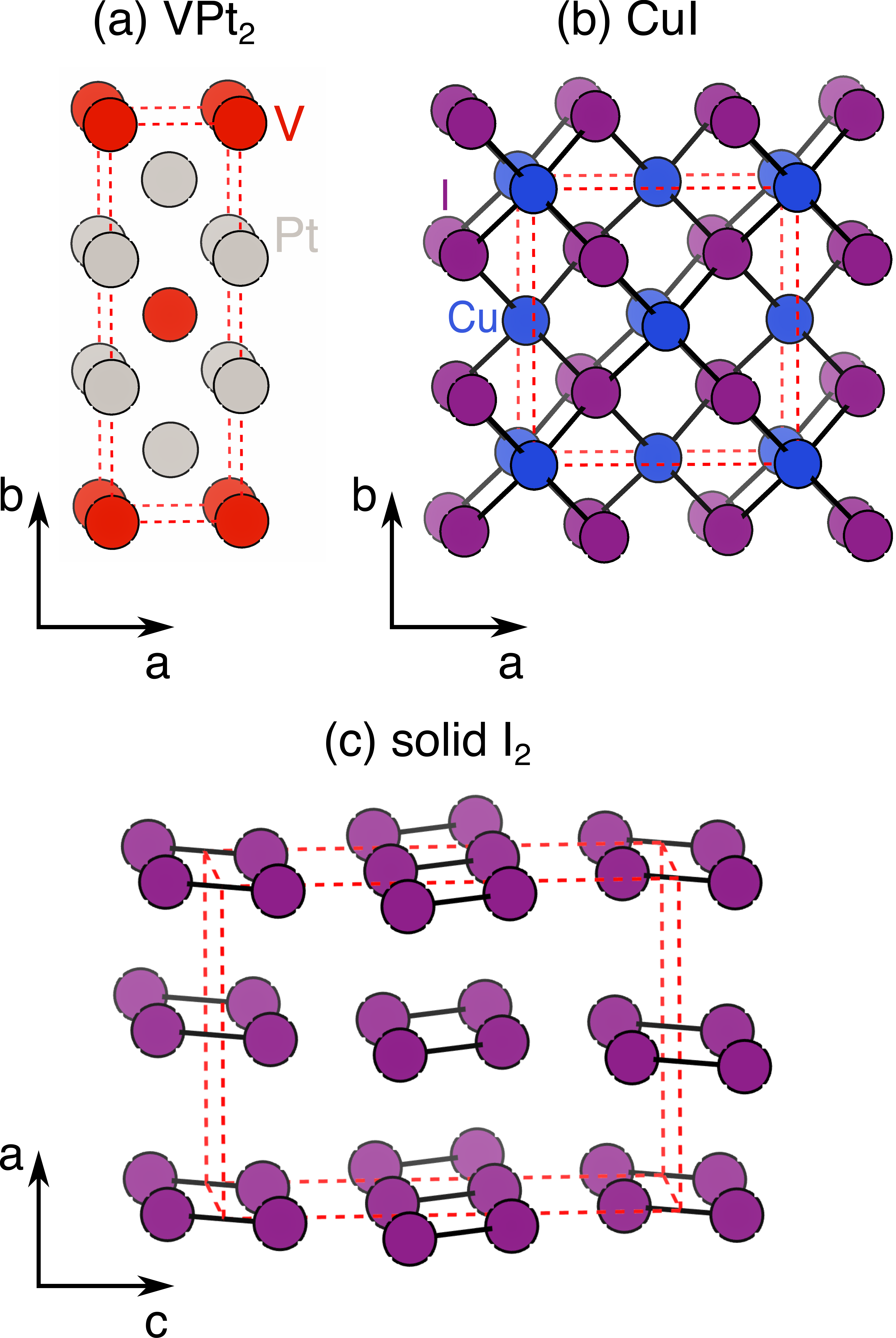}
  \caption{\label{fig:crystal_structures} Crystal structures of (a)
    VPt$_2$, (b) CuI, and (c) solid I$_2$. The red dashed lines
    indicate the conventional unit cell.}
\end{figure}

Here, we describe the systems of interest. The intermetallic VPt$_2$
crystallizes in the MoPt$_2$-type structure (orthorhombic, $Immm$
space group), shown in Fig. \ref{fig:crystal_structures}(a), which is
an ordered superstructure of the face-centered-cubic (fcc) structure
($Fm\bar{3}m$ space
group).\cite{giessen_crystal_1965,waterstrat_vanadium-platinum_1973,wolverton_ab_1993}
It becomes a disordered fcc solid solution above 1373
K.\cite{guo_standard_1994} VPt$_2$ is found to be a non-magnetic metal
in DFT.\cite{isaacs_performance_2018} The experimental $\Delta H_f$
value for VPt$_2$ of $-386 \pm 26$ meV/atom was obtained from direct
synthesis
calorimetry\cite{kim_experimental_2017,nash_two_2020,meschel_brief_2020}
by Guo and Kleppa.\cite{guo_standard_1994}

CuI is a superionic conductor with Cu sublattice
melting.\cite{funke_agi-type_1976} Below 642~K, it exists in the
zincblende crystal structure (cubic, $F\bar{4}3m$ space group, an
ordered diamond superstructure) shown in Fig.
\ref{fig:crystal_structures}(b).\cite{shan_description_2009} We
consider this phase, known as the $\gamma$ phase, in this work. It is
a $p$-type transparent semiconductor with a $\sim$ 3 eV band
gap.\cite{grundmann_cuprous_2013} Wurtzite ($\beta$) and
disordered-Cu$^+$ fcc ($\alpha$) phases of CuI emerge at higher
temperatures.\cite{miyake_phase_1952,rapoport_phase_1968} For CuI, an
assessed experimental value of $-369 \pm 10$ meV/atom comes from the
thermochemical database of the Russian Academy of Science and Moscow
State University.\cite{russian_database,belov2018ivtanthermo} This value is based on two
measurements: (1) heats of solution in FeCl$_3$ and HCl with a
Dewar-vessel isoperibol calorimeter at room
temperature\cite{cartwright_heats_1976} and (2) an equilibrium study
with elemental Cu and its amalgam.\cite{ishikawa_studies_1934}

Computing $\Delta H_f$ for the above compounds also requires
calculations for elemental vanadium, platinum, copper, and iodine. In
order to compare with experimental values, such calculations must correspond to
the standard reference states,\cite{dinsdale_sgte_1991} which are for
most elements the stable phase under standard conditions. The standard
reference states are the body-centered-cubic phase (bcc, $Im\bar{3}m$
space group) for V and the fcc phase for Pt and Cu, all non-magnetic
metals. As these are very simple and common structures, we do not
include them in Fig. \ref{fig:crystal_structures}. The standard
reference state for iodine is the solid phase with an orthorhombic
structure ($Cmce$ space group), a semiconductor with a band gap of 1.6
eV.\cite{yamamoto_ultraviolet_1987} Given that each iodine atom in this
structure has a single nearest neighbor at a distance of 2.7 \AA,
significantly smaller than the 2nd-nearest-neighbor distance of 3.5
\AA,\cite{van_bolhuis_refinement_1967} solid iodine can be considered
a molecular crystal (of I$_2$ molecules) and we refer to it as solid
I$_2$. However, we note that there is some evidence for a covalent,
rather than molecular, nature of the
crystal\cite{stepanov_iodine_2005}.

In bonded systems, the crystal field can act to suppress or magnify spin-orbit effects, depending on the space group of the solid or point group of a molecular system. For solids with high symmetry, {\em e.g.,} cubic systems, hybridization in one-particle states tends to average out a significant fraction of the spin-orbit effects. This applies even when the bonding patterns are not the same or even similar since the key point is breaking the lock of the atomic ground state symmetry and hybridizations in one-particle spinors. Unless there are some unusual symmetries that may enhance spin-orbit effects, we 
expect the effect of spin-orbit interactions on formation energies to be much smaller when compared to cohesive energy, since it involves the energy differences of solid energies only and the isolated atomic energies are irrelevant.

Since the phases relevant to
computing the $\Delta H_f$ values of interest are all solids in our
case, we ignore pressure-volume contributions and take the enthalpy to
be equal to the energy. In addition, the temperature dependence of the enthalpy is neglected, as is the role of any state of partial disorder in the compounds. 
For example, the formation energy of the CuI 
solid ($s$) can then be expressed in 
QMC total energies:
\begin{equation}
\label{eqn:H_via_total}
\Delta H_f({\rm CuI},s)=
E_{tot}({\rm CuI},s) -E_{tot}({\rm Cu},s)
-(1/2)E_{tot}({\rm I}_2, s)
\end{equation}
with all the quantities in Eq.  (\ref{eqn:H_via_total})
given per chemical formula as usual.\cite{crc_handbook} In what follows we further divide by the number of atoms in the formula of the target compound and therefore we report the values per atom.

\section{Computational Details}\label{sec:compdetails}

All calculations correspond to experimental  
lattice parameters and atomic positions 
for
VPt$_2$,\cite{giessen_crystal_1965} CuI,\cite{shan_description_2009},
bcc V,\cite{crc_handbook} fcc Pt,\cite{edwards_high_1951} fcc
Cu,\cite{davey_precision_1925} and solid
I$_2$.\cite{van_bolhuis_refinement_1967}

We employ scalar/full-relativistic, norm-conserving pseudopotentials to
describe the core electrons. For vanadium and copper, we use the
neon-core, local density approximation pseudopotentials of Krogel,
Santana, and Reboredo,\cite{krogel_pseudopotentials_2016} with the
optimized Rappe-Rabe-Kaxiras-Joannopoulos
form.\cite{rappe_optimized_1990} For platinum, we use a PBE
18-valence-electron, Troullier-Martins\cite{troullier_efficient_1991}
pseudopotential. For iodine, we use the
Burkatzki-Filippi-Dolg Hartree-Fock
pseudopotential\cite{burkatzki_energy-consistent_2007} with $5s$ and
$5p$ in the valence. 
For spin-orbit calculations, we adopt pseudopotential spin-orbit terms from the Stuttgart group for I\cite{STU} and Pt\cite{STU_Pt} elements, while for the Cu and V elements spin-orbit effects are neglected.
We include validation of the iodine pseudopotential based on atomic ionization potential and electron affinity calculations as well as SO multiplet splittings in Appendix D.

Single-particle wavefunctions from DFT within the PBE
exchange-correlation DFA were generated using the
\textsc{quantum espresso} package.\cite{giannozzi_advanced_2017} We
employ a wavefunction plane wave kinetic energy cutoff of 350 Ry for
VPt$_2$, Pt, and V and 450 Ry for CuI, Cu, and solid I$_2$. The Brillouin zone was sampled
with uniform $k$-grids of at least 500 $k$-points/\AA$^{-3}$.
Using these parameters provided one-particle orbitals with accuracy appropriate for use in QMC methods.
PBE0+spin-orbit calculations were also carried out using \textsc{quantum espresso} with the same kinetic energy cutoffs as above.

QMC calculations with the fixed-node diffusion Monte Carlo (DMC)
method were performed with \textsc{qmcpack}.\cite{QMCPACK} We use the
Slater-Jastrow form of the trial many-body wavefunction with optimized
one-, two-, and three-body Jastrow correlation
factors.\cite{jastrow_many-body_1955,drummond_jastrow_2004} A fixed
0.005 Ha$^{-1}$ time step was employed. We use two finite size
corrections: the model periodic Coulomb interaction
(MPC)\cite{williamson_elimination_1997,kent_finite-size_1999,drummond_finite-size_2008}
and the Chiesa-Ceperley-Martin-Holzmann kinetic energy
correction.\cite{chiesa_finite-size_2006,drummond_finite-size_2008} We
employ twist-averaged boundary
conditions\cite{lin_twist-averaged_2001} to remove one-body finite
size effects and finite size extrapolation of supercell results to
remove two-body finite size effects. Optimal supercells were generated
via \textsc{nexus}\cite{krogel_nexus_2016}.
Further data on the QMC calculations can be found 
in the Appendices.

We also performed DFT calculations of the spin-orbit corrections to the I$_2$ dimer binding energy, the CuI dimer binding energy, and solid CuI using the FHI-aims code\cite{blum2009ab,ren2012,marek2014elpa} with the PBE0 exchange-correlation functional\cite{PBE0} and the default tight basis sets. For the CuI dimer, we first optimized the dimer bond length so that forces were less than 0.5~meV/{\AA}.
Those results did not differ in any  discernible way from the results using \textsc{quantum espresso}.

Another set of calculations was carried out using the \textsc{dirac} code.\cite{DIRAC19, saue_dirac_2020}
Specifically, we evaluated total energy shifts going from the spin-averaged case (AREP) to the full spin-orbit case (SOREP) for the I$_2$ and CuI molecules using the aug-cc-pVTZ basis set.
I$_2$ molecular energies were calculated using the CCSD(T) method, while the COSCI (complete open-shell CI) method was used for the CuI molecule. 

Considering the spin-orbit corrections, we write the formation energy as a sum of spin-averaged (AREP) contribution and the spin-orbit correction, for example, for CuI we have:
\begin{equation}
\Delta H_f({\rm CuI},s) = 
\Delta H_f({\rm CuI},s)_{\rm AREP} + \Delta^{SO}.
\end{equation}
There are only two contributions to $\Delta^{SO}$
\begin{equation}
\label{eqn:so_terms}
\Delta^{SO}=\Delta^{SO}({\rm CuI},s)-(1/2)\Delta^{SO}({\rm I}_2,s)   
\end{equation}
since we assume that the corresponding value from Cu is negligible. Further details are discussed in the next section
and in the Appendices.

As explicit spin-orbit calculations using the DMC method are prohibitively expensive for very large supercell calculations used in this work, we employ the PBE0+spin-orbit contribution $\Delta^{SO}$ for DMC values as well.
We believe this is justified since PBE0 values without spin-orbit effects agree very well with DMC results as will be shown later.
Note that there are significant recent efforts to evaluate spin-orbit effects using QMC in moderately large systems \cite{chang_effective_2020, annaberdiyev_electronic_2022}. 

\section{Results and Discussion}\label{sec:results}

\begin{figure*}[htbp]
  \centering
  \includegraphics[width=\textwidth]{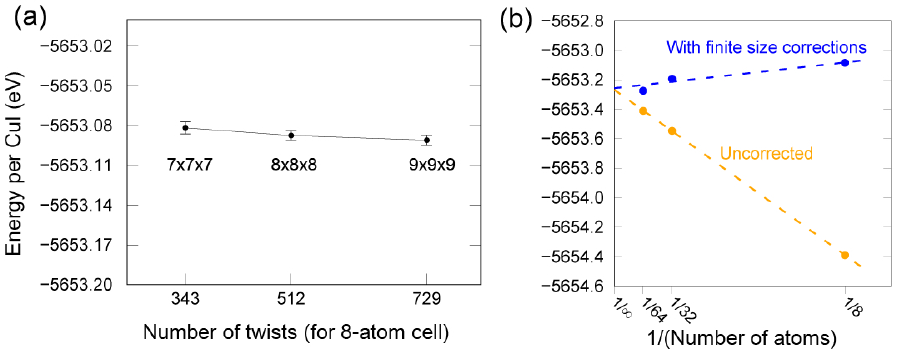}
  \caption{\label{fig:convergence} (a) Convergence of energy with
    respect to twist mesh and (b) finite size extrapolation of
    energies with and without finite size corrections for CuI.
    }
\end{figure*}

Due to very large total energies from semicores ($3s^23p^6$) of transition elements, small to medium size supercells (4--16 atoms) are employed to assess the convergence
of the energy with respect to twist mesh. Let us consider the CuI case, for
which we use the eight-atom conventional cell, as an example. As shown in
Fig. \ref{fig:convergence}(a), the energy is well converged for a
twist mesh of 8$\times$8$\times$8.

In order to address two-body finite-size effects, we perform
calculations for several supercell sizes, using twist meshes
corresponding to the same converged twist density. Fig.
\ref{fig:convergence}(b) shows, taking CuI again as an example, the extrapolation of the energies of different supercells to the
thermodynamic (bulk) limit. Notably, the data are approximately linear
in the inverse number of atoms, and extrapolations of the uncorrected
and corrected (i.e., using the MPC and kinetic energy corrections)
data yield nearly identical values (difference of 11 meV per formula
unit). We note that we find a negligible need for finite-size extrapolation
for solid I$_2$ when finite-size corrections are employed (as shown in the Appendix C). This
finding that the interactions are relatively short-ranged is
consistent with solid iodine's nature as a molecular crystal; similar
behavior was found previously for molecular crystals of NH$_3$ and
CO$_2$\cite{zen_fast_2018}.

\begin{figure*}[htbp]
  \centering
  \includegraphics[width=0.70\linewidth]{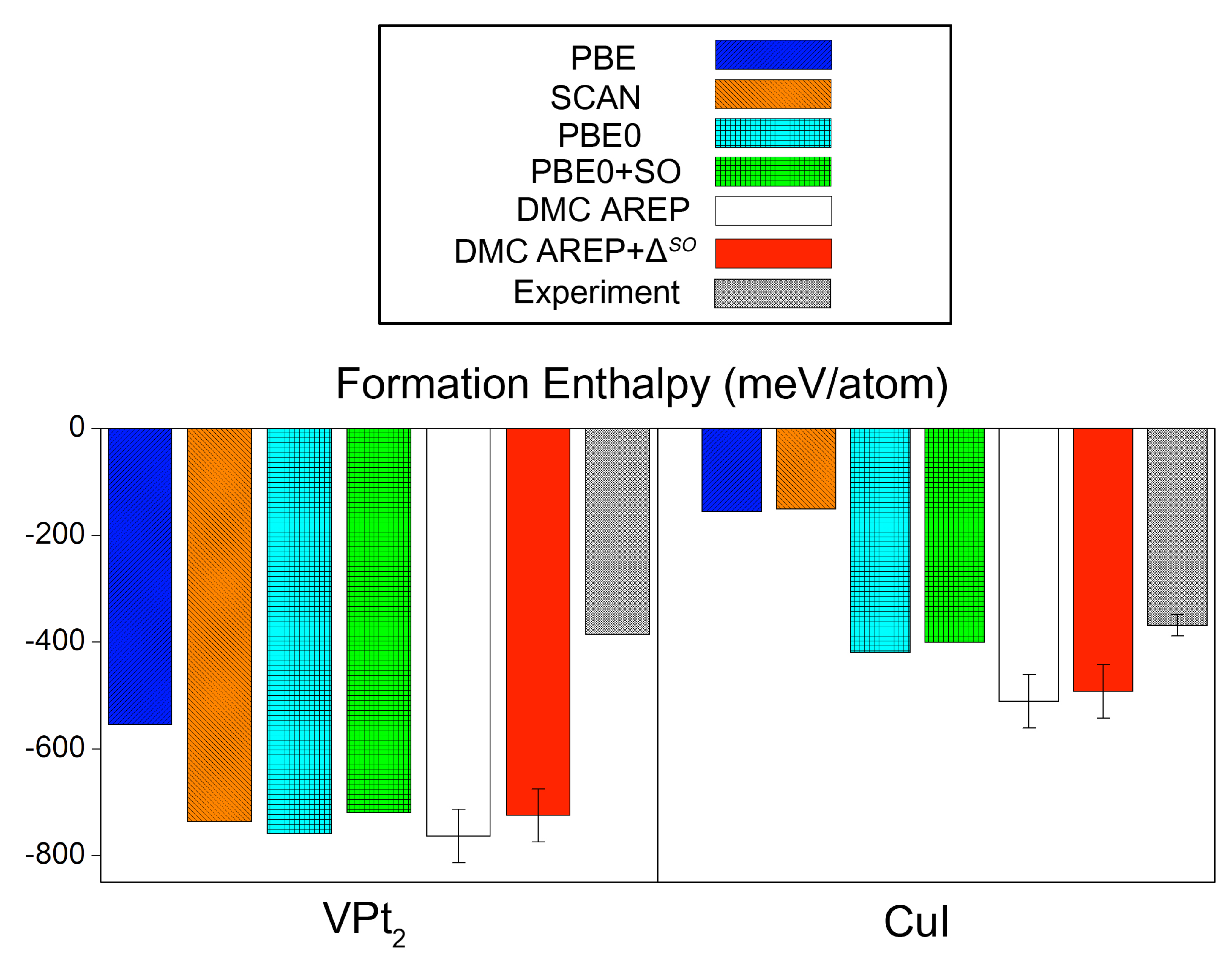}
  \caption{\label{fig:fe} Computed and experimental compound formation
    enthalpy of VPt$_2$ (left) and CuI (right).
    }
\end{figure*}

Figure~\ref{fig:fe} contains the main result of this work: QMC values of $\Delta H_f$ for VPt$_2$ and CuI compared with other methods and experiments. The PBE and SCAN results
shown for comparison are taken from past work using the projector
augmented wave method in the \textsc{vasp}
code\cite{isaacs_performance_2018}. $\Delta H_f$ computed from our own
PBE calculations here (performed to generate trial wavefunctions)
agree with the literature values to within $\sim$ 30 meV/atom,
confirming the DFT-predicted $\Delta H_f$ are not terribly sensitive
to the choice of DFT code, pseudopotentials, and other computational
details.
PBE0 and PBE0+SO values are results from this work.

The $\Delta H_f$ for VPt$_2$ from QMC calculations
without spin-orbit is  $-764\pm7$~meV/atom with the nominal statistical error which is very small due to the fact that it is an intensive quantity.
Note however, that we do not consider this statistical uncertainty as being really representative, say, for comparisons with the DFT values. The overall error generated by the key systematic bias, which is the fixed-node
approximation from the single-reference trial function, is likely much larger. 
Considering that energy differences involve partial fixed-node error cancellation and per atom normalization, we assume a systematic error at the level of $\approx$ 50~meV/atom. In order to quantitatively ascertain this level, we would need multi-reference trial functions data with significantly lower fixed-node bias, which unfortunately are not available at present. Therefore, we use a semi-quantitative guess based on recent QMC calculations such as crystal of LaScO$_3$ \cite{melton2020},
solid Si with up to 216
atoms in the supercell \cite{annaberdiyev_cohesion_2021} and also estimations of exact correlation energies for $3d$ atoms \cite{gani2020}. In what follows, we therefore assume this value as an approximate error bar which includes both random uncertainty and systematic biases with the caveats mentioned above.

Comparison with DFT results reveals that
 the  formation enthalpy for VPt$_2$ calculated within the SCAN DFA  ($-737$~meV/atom) 
is very close to the QMC value ($-764\pm50$~meV/atom) whereas the PBE-predicted formation enthalpy ($-555$~meV/atom) is much
too small in magnitude (by 209 meV/atom).
It is worth noting that the PBE0 result ($-759$~meV/atom),
which completes the overall picture with a hybrid DFA, is in the same range
as SCAN and  QMC.
In a previous larger-scale
benchmark, SCAN was found to offer no improvement from PBE in
predicting $\Delta H_f$ of ``weakly-bound'' compounds (experimental
$|\Delta H_f| < 1$ eV/atom), which are mainly intermetallics like
VPt$_2$.\cite{isaacs_performance_2018} In fact, the mean absolute
error compared to experiment for such compounds was found to be 20\%
larger for SCAN than for PBE. This observation is surprising from the
perspective that intermetallics should be relatively easy to describe
given (1) they are metals with delocalized electronic states and (2)
in the metallic regime (of orbital kinetic energy density), SCAN is
constructed to accurately describe slowly-varying densities, just as
PBE is. Although it pertains to only a single compound, which could be an exception, the VPt$_2$
result here  suggests that the SCAN accuracy probably varies even if we consider just one class of compounds such as intermetallics.
The $\Delta^{SO}$ contribution to VPt$_2$ calculated using PBE0+SO is 39~meV/atom; 
therefore, it does not change the overall conclusion since it is only $\approx$ 5\% of the calculated formation value, see Fig. \ref{fig:fe}.

Our results raise the important
question of why the experimental $\Delta H_f$ value for VPt$_2$
deviates so significantly from theory. One contributing factor may be
the incomplete nature of the synthesis reaction in the measurement. As
noted by Guo and Kleppa, multiple experimental probes (X-ray
diffraction, scanning electron microscopy, and energy-dispersive
microanalysis) showed the presence of vanadium oxide with the
VPt$_2$.\cite{guo_standard_1994} Although the amount observed was
small (a few percent), this may contribute to an appreciable
underestimation of the $\Delta H_f$ magnitude, especially due to the
large $\Delta H_f$ for vanadium oxide.

We note that the case of VPt$_3$ appears to be very similar: vanadium
oxide was found in experimental samples, and the measured $\Delta H_f$ value of
$-284 \pm 19$ meV/atom is much smaller in magnitude than that of PBE
($-457$ meV/atom) and SCAN ($-603$
meV/atom).\cite{guo_standard_1994,isaacs_performance_2018}
Interestingly, the enhancement in $\Delta H_f$ in SCAN as compared to
PBE is almost the same (factor of $1.32-1.33$) for both cases. Our
results should motivate a re-investigation of the experimental $\Delta H_f$
for these compounds. More generally, we speculate that errors in
experimental $\Delta H_f$ may also help explain other recent work in
which experimental $\Delta H_f$ were found to differ from high-level
(namely, random phase approximation) calculation
results.\cite{nepal_formation_2020}

The CuI results stand in clear contrast to those of VPt$_2$. Here, the
PBE and SCAN $\Delta H_f$ values are nearly identical ($\approx -153$
meV/atom) and are much too small in magnitude compared to the experimental value. The QMC result without including spin-orbit corrections ($-511$ meV/atom) is significantly closer
to the experimental value but still is appreciably different from it. Therefore, we consider spin-orbit interactions as a possible source of the remaining discrepancy.

In order to evaluate the impact of spin-orbit on CuI, we use 
Eq.~(\ref{eqn:so_terms}). 
As discussed previously, for this we use the PBE0 functional that gives the closest agreement with QMC at the AREP (averaged spin-orbit) level.
Based on this, we estimate the effect of spin-orbit explicitly within the PBE0 DFA framework. The effect goes in the direction of becoming closer to the experiment, however, the correction is rather small, see, Fig. \ref{fig:fe}.  
We probe this correction also in an alternative way using CuI and I$_2$ molecular energy shifts that involve explicit spin-orbit COSCI, CCSD(T) calculations as well as DFT differences which gives similar results (see  Appendix D).
The resulting correction is given by:
\begin{equation}
\Delta^{SO}=\Delta^{SO}({\rm CuI},s)- (1/2)\Delta^{SO}({\rm I}_2,s)= 18\; {\rm meV/atom.}
\end{equation}
This correction reduces the QMC AREP formation energy of $-$511~meV/atom to 
$-$493~meV/atom which is closer to the experimental value of  $-369\pm20$~meV/atom where we assume an effective error bar per differences between independent experiments. 
Therefore, 
incorporating spin-orbit corrections for CuI and solid I$_2$ brings theory and experiment into an imperfect but improved agreement within $\sim$ 120~meV/atom.

\begin{figure*}[htbp]
  \centering
  \includegraphics[width=6.5 truein]{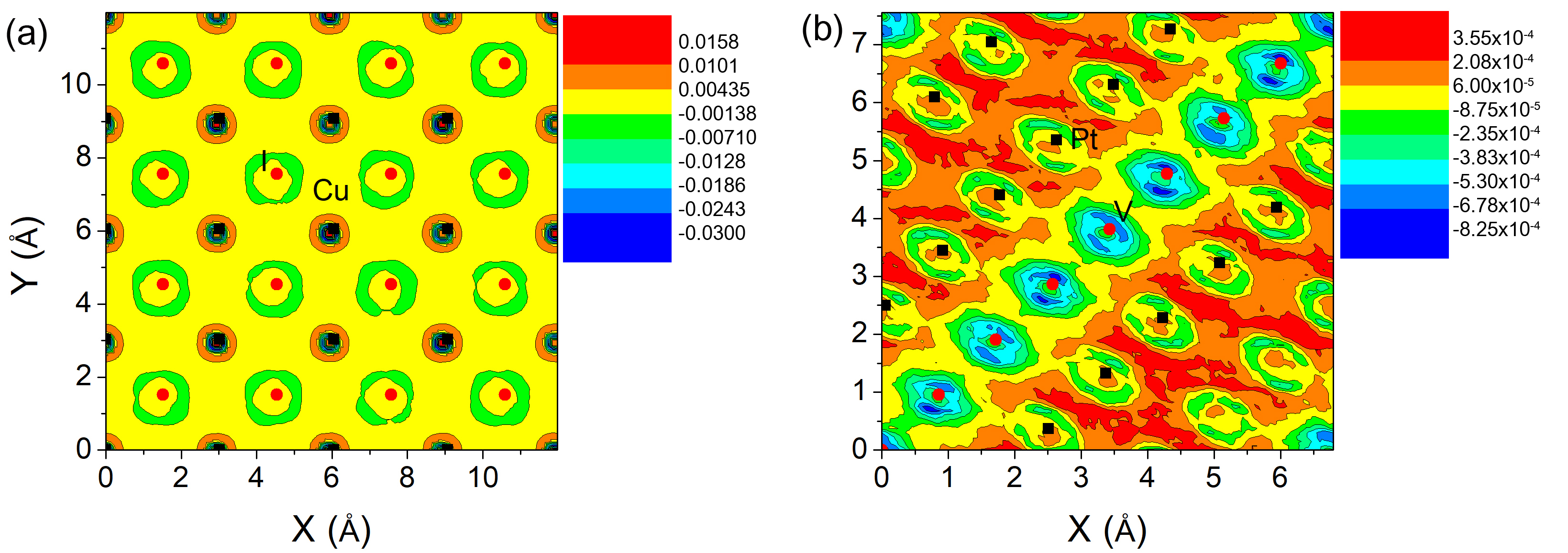}
  \caption{\label{fig:den} Total density difference between DMC and PBE for (a) CuI and (b) VPt$_2$ projected onto the (001) plane.}
  \label{fig:density}
\end{figure*}  

Apart from the mild spin-orbit shift, the exchange-correlation description not only in PBE but also within the SCAN DFA appears to be of mixed quality. Given the known difficulty of describing
localized $3d$ transition metal states, which give rise to significant self-interaction effects especially in insulators, we believe the presence of Cu
$3d$ orbitals may be responsible for the large PBE and SCAN errors: Vanadium $d$ states are nominally singly occupied while Cu $d$-states are more localized and almost fully doubly occupied with more pronounced Hubbard physics. This is also supported by the PBE0 result above that enhances the localization of $3d$ levels in the like-spin channel and therefore effectively increases the repulsion in the unlike-spin channel. 
In line with this view, we note that beyond-DFT Hubbard $U$
corrections were found to be necessary even with SCAN in describing
gapped Li$_x$FePO$_4$ and Li$_x$Mn$_2$O$_4$ systems, which also
contain $3d$ transition metal states.\cite{isaacs_prediction_2020} It is instructive to illustrate this by examining the density differences between DMC and DFT (PBE). Figure~\ref{fig:density} shows the CuI (left panel) and VPt$_2$ density differences projected onto the (001) plane. The left panel shows that the density differences are well localized around Cu and I sites, with the difference being slightly negative (DMC density smaller than DFT) near I sites, and positive (DMC density larger than DFT) near Cu sites, indicating a difference in the description of densities arising from orbitals near the Cu sites. In contrast, the VPt$_2$ density differences are significantly smaller in magnitude and more diffuse in the (001) plane.

\section{Conclusions}\label{sec:conclusions}
We employed QMC to calculate the formation energies for intermetallic VPt$_2$ and semiconducting CuI and compared to DFT within the PBE and SCAN DFA, hybrid PBE0 DFA DFT, and experiment. 
QMC calculations have been carried out with single-reference trial functions and as such, they correspond to the level of accuracy that we colloquially refer to as the QMC ``standard model". 
It has been applied to a plethora of systems including solids with supercells containing hundreds of valence
electrons with rather consistent improvements and insights over more traditional approaches. In this study, we attempt to advance this approach for the resulting very small energy differences that are likely at the edge of its accuracy limits. We note that the computational demands
are substantial, especially if we consider that 
the total energy of a single Cu atom with semi-core  $3s^23p^6$
in valence space is above  5000~eV so an accuracy of a few tens of meV is very challenging to achieve. 

For VPt$_2$, we find  close agreement between QMC, SCAN, and PBE0 results. 
On the other hand, the difference with the experimental value is significantly larger, suggesting a need for an independent new measurement that would confirm or update the currently available data.
For CuI, we find that the QMC formation energy is in a rough  agreement with the experimental value within approximately 120  meV/atom, after adding corrections that take into account spin-orbit interactions in CuI and solid I$_2$. 
In our estimations, the spin-orbit effects from iodine appear rather small in both DFT using the PBE0 functional and in estimations using COSCI and CCSD(T) that involved CuI and I$_2$ molecules using two-component spinors.
The large difference between the DFT within the PBE and SCAN DFA and QMC likely arises from difficulties in the description of localized Cu $3d$ orbitals in CuI, consistent with large electron density differences between QMC and DFT-PBE near the Cu sites [Fig.~\ref{fig:density} (a)]. Our results indicate that, for systems like CuI, significant exchange-correlation errors are still present even in the SCAN DFA so they likely limit its predictive power. 

This study is the first attempt to address more weakly bonded  intermetallics with heavy elements by QMC methods.
Clearly, in order to pinpoint imperfections with higher energy resolution and to better control fixed-node bias, new developments with multi-reference trial functions and accurate pseudopotentials are highly desirable. 
Since not much has been known
about such systems using many-body wave function methods so far, we consider our results very encouraging and promising for future studies.

\section{Data Availability}
See Appendices for tabulated formation energies, pseudopotential tests, finite-size extrapolation plots, additional estimation of spin-orbit effect on CuI formation energy, CuI DFT band gaps, QMC variances, and other supporting data.
Input and output files generated in this work are available in Materials Data Facility \cite{blaiszik_materials_2016, blaiszik_data_2019} and can be found at \cite{mdf_data}.

\bigskip

\begin{acknowledgments}
  We acknowledge support from the U.S. Department of Energy, Office of
  Science, Basic Energy Sciences, Materials Sciences and Engineering
  Division, as part of the Computational Materials Sciences Program
  and Center for Predictive Simulation of Functional Materials. An
  award of computer time was provided by the Innovative and Novel
  Computational Impact on Theory and Experiment (INCITE) program. This
  research used resources of the Argonne Leadership Computing
  Facility, which is a DOE Office of Science User Facility supported
  under contract DE-AC02-06CH11357 and resources of the Oak Ridge
  Leadership Computing Facility, which is a DOE Office of Science User
  Facility supported under Contract DE-AC05-00OR22725. 
  
  We thank Vinay
  Hegde and James Saal (Citrine Informatics), Philip Nash (Illinois
  Institute of Technology), and Paul Kent and Jaron Krogel (ORNL) for
  useful discussions. We gratefully acknowledge the computing resources provided on Bebop and Blues, high-performance computing clusters operated by the Laboratory Computing Resource Center at Argonne National Laboratory.
  
  This research used resources of the National Energy Research Scientific Computing Center (NERSC), a U.S. Department of Energy Office of Science User Facility located at Lawrence Berkeley National Laboratory, operated under Contract No. DE-AC02-05CH11231.
  
Part of this research (DIRAC and DFT/PBE0 calculations) was conducted by A.A. at the Center for Nanophase Materials Sciences (CNMS), which is a DOE Office of Science User Facility.
  
  This manuscript has been authored by UT-Battelle, LLC under Contract No. DE-AC05-00OR22725 with the U.S. Department of Energy. The United States Government retains and the publisher, by accepting the article for publication, acknowledges that the United States Government retains a non-exclusive, paid-up, irrevocable, world-wide license to publish or reproduce the published form of this manuscript, or allow others to do so, for United States Government purposes. The Department of Energy will provide public access to these results of federally sponsored research in accordance with the DOE Public Access Plan (http://energy.gov/downloads/doe-public-access-plan).

\end{acknowledgments}

\bibliography{main}

\begin{thebibliography}{84}%
\makeatletter
\providecommand \@ifxundefined [1]{%
 \@ifx{#1\undefined}
}%
\providecommand \@ifnum [1]{%
 \ifnum #1\expandafter \@firstoftwo
 \else \expandafter \@secondoftwo
 \fi
}%
\providecommand \@ifx [1]{%
 \ifx #1\expandafter \@firstoftwo
 \else \expandafter \@secondoftwo
 \fi
}%
\providecommand \natexlab [1]{#1}%
\providecommand \enquote  [1]{``#1''}%
\providecommand \bibnamefont  [1]{#1}%
\providecommand \bibfnamefont [1]{#1}%
\providecommand \citenamefont [1]{#1}%
\providecommand \href@noop [0]{\@secondoftwo}%
\providecommand \href [0]{\begingroup \@sanitize@url \@href}%
\providecommand \@href[1]{\@@startlink{#1}\@@href}%
\providecommand \@@href[1]{\endgroup#1\@@endlink}%
\providecommand \@sanitize@url [0]{\catcode `\\12\catcode `\$12\catcode
  `\&12\catcode `\#12\catcode `\^12\catcode `\_12\catcode `\%12\relax}%
\providecommand \@@startlink[1]{}%
\providecommand \@@endlink[0]{}%
\providecommand \url  [0]{\begingroup\@sanitize@url \@url }%
\providecommand \@url [1]{\endgroup\@href {#1}{\urlprefix }}%
\providecommand \urlprefix  [0]{URL }%
\providecommand \Eprint [0]{\href }%
\providecommand \doibase [0]{http://dx.doi.org/}%
\providecommand \selectlanguage [0]{\@gobble}%
\providecommand \bibinfo  [0]{\@secondoftwo}%
\providecommand \bibfield  [0]{\@secondoftwo}%
\providecommand \translation [1]{[#1]}%
\providecommand \BibitemOpen [0]{}%
\providecommand \bibitemStop [0]{}%
\providecommand \bibitemNoStop [0]{.\EOS\space}%
\providecommand \EOS [0]{\spacefactor3000\relax}%
\providecommand \BibitemShut  [1]{\csname bibitem#1\endcsname}%
\let\auto@bib@innerbib\@empty
\bibitem [{\citenamefont {Bartel}\ \emph {et~al.}(2020)\citenamefont {Bartel},
  \citenamefont {Trewartha}, \citenamefont {Wang}, \citenamefont {Dunn},
  \citenamefont {Jain},\ and\ \citenamefont {Ceder}}]{bartel_critical_2020}%
  \BibitemOpen
  \bibfield  {author} {\bibinfo {author} {\bibfnamefont {C.~J.}\ \bibnamefont
  {Bartel}}, \bibinfo {author} {\bibfnamefont {A.}~\bibnamefont {Trewartha}},
  \bibinfo {author} {\bibfnamefont {Q.}~\bibnamefont {Wang}}, \bibinfo {author}
  {\bibfnamefont {A.}~\bibnamefont {Dunn}}, \bibinfo {author} {\bibfnamefont
  {A.}~\bibnamefont {Jain}}, \ and\ \bibinfo {author} {\bibfnamefont
  {G.}~\bibnamefont {Ceder}},\ }\href {\doibase 10.1038/s41524-020-00362-y}
  {\bibfield  {journal} {\bibinfo  {journal} {npj Comput. Mater.}\ }\textbf
  {\bibinfo {volume} {6}},\ \bibinfo {pages} {1} (\bibinfo {year}
  {2020})}\BibitemShut {NoStop}%
\bibitem [{\citenamefont {Hohenberg}\ and\ \citenamefont
  {Kohn}(1964)}]{hohenberg_inhomogeneous_1964}%
  \BibitemOpen
  \bibfield  {author} {\bibinfo {author} {\bibfnamefont {P.}~\bibnamefont
  {Hohenberg}}\ and\ \bibinfo {author} {\bibfnamefont {W.}~\bibnamefont
  {Kohn}},\ }\href@noop {} {\bibfield  {journal} {\bibinfo  {journal} {Phys.
  Rev.}\ }\textbf {\bibinfo {volume} {136}},\ \bibinfo {pages} {B864} (\bibinfo
  {year} {1964})}\BibitemShut {NoStop}%
\bibitem [{\citenamefont {Kohn}\ and\ \citenamefont
  {Sham}(1965)}]{kohn_self-consistent_1965}%
  \BibitemOpen
  \bibfield  {author} {\bibinfo {author} {\bibfnamefont {W.}~\bibnamefont
  {Kohn}}\ and\ \bibinfo {author} {\bibfnamefont {L.~J.}\ \bibnamefont
  {Sham}},\ }\href@noop {} {\bibfield  {journal} {\bibinfo  {journal} {Phys.
  Rev.}\ }\textbf {\bibinfo {volume} {140}},\ \bibinfo {pages} {A1133}
  (\bibinfo {year} {1965})}\BibitemShut {NoStop}%
\bibitem [{\citenamefont {Jones}(2015)}]{jones_density_2015}%
  \BibitemOpen
  \bibfield  {author} {\bibinfo {author} {\bibfnamefont {R.}~\bibnamefont
  {Jones}},\ }\href {\doibase 10.1103/RevModPhys.87.897} {\bibfield  {journal}
  {\bibinfo  {journal} {Rev. Mod. Phys.}\ }\textbf {\bibinfo {volume} {87}},\
  \bibinfo {pages} {897} (\bibinfo {year} {2015})}\BibitemShut {NoStop}%
\bibitem [{\citenamefont {Marzari}\ \emph {et~al.}(2021)\citenamefont
  {Marzari}, \citenamefont {Ferretti},\ and\ \citenamefont
  {Wolverton}}]{marzari2021electronic}%
  \BibitemOpen
  \bibfield  {author} {\bibinfo {author} {\bibfnamefont {N.}~\bibnamefont
  {Marzari}}, \bibinfo {author} {\bibfnamefont {A.}~\bibnamefont {Ferretti}}, \
  and\ \bibinfo {author} {\bibfnamefont {C.}~\bibnamefont {Wolverton}},\
  }\href@noop {} {\bibfield  {journal} {\bibinfo  {journal} {Nature materials}\
  }\textbf {\bibinfo {volume} {20}},\ \bibinfo {pages} {736} (\bibinfo {year}
  {2021})}\BibitemShut {NoStop}%
\bibitem [{\citenamefont {Stevanovic}\ \emph {et~al.}(2012)\citenamefont
  {Stevanovic}, \citenamefont {Lany}, \citenamefont {Zhang},\ and\
  \citenamefont {Zunger}}]{stevanovic_correcting_2012}%
  \BibitemOpen
  \bibfield  {author} {\bibinfo {author} {\bibfnamefont {V.}~\bibnamefont
  {Stevanovic}}, \bibinfo {author} {\bibfnamefont {S.}~\bibnamefont {Lany}},
  \bibinfo {author} {\bibfnamefont {X.}~\bibnamefont {Zhang}}, \ and\ \bibinfo
  {author} {\bibfnamefont {A.}~\bibnamefont {Zunger}},\ }\href {\doibase
  10.1103/PhysRevB.85.115104} {\bibfield  {journal} {\bibinfo  {journal} {Phys.
  Rev. B}\ }\textbf {\bibinfo {volume} {85}},\ \bibinfo {pages} {115104}
  (\bibinfo {year} {2012})}\BibitemShut {NoStop}%
\bibitem [{\citenamefont {Kirklin}\ \emph {et~al.}(2015)\citenamefont
  {Kirklin}, \citenamefont {Saal}, \citenamefont {Meredig}, \citenamefont
  {Thompson}, \citenamefont {Doak}, \citenamefont {Aykol}, \citenamefont
  {R\"{u}hl},\ and\ \citenamefont {Wolverton}}]{kirklin_open_2015}%
  \BibitemOpen
  \bibfield  {author} {\bibinfo {author} {\bibfnamefont {S.}~\bibnamefont
  {Kirklin}}, \bibinfo {author} {\bibfnamefont {J.~E.}\ \bibnamefont {Saal}},
  \bibinfo {author} {\bibfnamefont {B.}~\bibnamefont {Meredig}}, \bibinfo
  {author} {\bibfnamefont {A.}~\bibnamefont {Thompson}}, \bibinfo {author}
  {\bibfnamefont {J.~W.}\ \bibnamefont {Doak}}, \bibinfo {author}
  {\bibfnamefont {M.}~\bibnamefont {Aykol}}, \bibinfo {author} {\bibfnamefont
  {S.}~\bibnamefont {R\"{u}hl}}, \ and\ \bibinfo {author} {\bibfnamefont
  {C.}~\bibnamefont {Wolverton}},\ }\href {\doibase
  10.1038/npjcompumats.2015.10} {\bibfield  {journal} {\bibinfo  {journal} {npj
  Comput. Mater.}\ }\textbf {\bibinfo {volume} {1}},\ \bibinfo {pages} {15010}
  (\bibinfo {year} {2015})}\BibitemShut {NoStop}%
\bibitem [{\citenamefont {Zhang}\ \emph {et~al.}(2018)\citenamefont {Zhang},
  \citenamefont {Kitchaev}, \citenamefont {Yang}, \citenamefont {Chen},
  \citenamefont {Dacek}, \citenamefont {Sarmiento-Pérez}, \citenamefont
  {Marques}, \citenamefont {Peng}, \citenamefont {Ceder}, \citenamefont
  {Perdew},\ and\ \citenamefont {Sun}}]{zhang_efficient_2018}%
  \BibitemOpen
  \bibfield  {author} {\bibinfo {author} {\bibfnamefont {Y.}~\bibnamefont
  {Zhang}}, \bibinfo {author} {\bibfnamefont {D.~A.}\ \bibnamefont {Kitchaev}},
  \bibinfo {author} {\bibfnamefont {J.}~\bibnamefont {Yang}}, \bibinfo {author}
  {\bibfnamefont {T.}~\bibnamefont {Chen}}, \bibinfo {author} {\bibfnamefont
  {S.~T.}\ \bibnamefont {Dacek}}, \bibinfo {author} {\bibfnamefont {R.~A.}\
  \bibnamefont {Sarmiento-Pérez}}, \bibinfo {author} {\bibfnamefont
  {M.~A.~L.}\ \bibnamefont {Marques}}, \bibinfo {author} {\bibfnamefont
  {H.}~\bibnamefont {Peng}}, \bibinfo {author} {\bibfnamefont {G.}~\bibnamefont
  {Ceder}}, \bibinfo {author} {\bibfnamefont {J.~P.}\ \bibnamefont {Perdew}}, \
  and\ \bibinfo {author} {\bibfnamefont {J.}~\bibnamefont {Sun}},\ }\href
  {\doibase 10.1038/s41524-018-0065-z} {\bibfield  {journal} {\bibinfo
  {journal} {npj Comput. Mater.}\ }\textbf {\bibinfo {volume} {4}},\ \bibinfo
  {pages} {9} (\bibinfo {year} {2018})}\BibitemShut {NoStop}%
\bibitem [{\citenamefont {Bartel}\ \emph {et~al.}(2019)\citenamefont {Bartel},
  \citenamefont {Weimer}, \citenamefont {Lany}, \citenamefont {Musgrave},\ and\
  \citenamefont {Holder}}]{bartel_role_2019}%
  \BibitemOpen
  \bibfield  {author} {\bibinfo {author} {\bibfnamefont {C.~J.}\ \bibnamefont
  {Bartel}}, \bibinfo {author} {\bibfnamefont {A.~W.}\ \bibnamefont {Weimer}},
  \bibinfo {author} {\bibfnamefont {S.}~\bibnamefont {Lany}}, \bibinfo {author}
  {\bibfnamefont {C.~B.}\ \bibnamefont {Musgrave}}, \ and\ \bibinfo {author}
  {\bibfnamefont {A.~M.}\ \bibnamefont {Holder}},\ }\href {\doibase
  10.1038/s41524-018-0143-2} {\bibfield  {journal} {\bibinfo  {journal} {npj
  Comput. Mater.}\ }\textbf {\bibinfo {volume} {5}},\ \bibinfo {pages} {4}
  (\bibinfo {year} {2019})}\BibitemShut {NoStop}%
\bibitem [{\citenamefont {Friedrich}\ \emph {et~al.}(2019)\citenamefont
  {Friedrich}, \citenamefont {Usanmaz}, \citenamefont {Oses}, \citenamefont
  {Supka}, \citenamefont {Fornari}, \citenamefont {Buongiorno~Nardelli},
  \citenamefont {Toher},\ and\ \citenamefont
  {Curtarolo}}]{friedrich_coordination_2019}%
  \BibitemOpen
  \bibfield  {author} {\bibinfo {author} {\bibfnamefont {R.}~\bibnamefont
  {Friedrich}}, \bibinfo {author} {\bibfnamefont {D.}~\bibnamefont {Usanmaz}},
  \bibinfo {author} {\bibfnamefont {C.}~\bibnamefont {Oses}}, \bibinfo {author}
  {\bibfnamefont {A.}~\bibnamefont {Supka}}, \bibinfo {author} {\bibfnamefont
  {M.}~\bibnamefont {Fornari}}, \bibinfo {author} {\bibfnamefont
  {M.}~\bibnamefont {Buongiorno~Nardelli}}, \bibinfo {author} {\bibfnamefont
  {C.}~\bibnamefont {Toher}}, \ and\ \bibinfo {author} {\bibfnamefont
  {S.}~\bibnamefont {Curtarolo}},\ }\href {\doibase 10.1038/s41524-019-0192-1}
  {\bibfield  {journal} {\bibinfo  {journal} {npj Comput. Mater.}\ }\textbf
  {\bibinfo {volume} {5}},\ \bibinfo {pages} {1} (\bibinfo {year}
  {2019})}\BibitemShut {NoStop}%
\bibitem [{\citenamefont {Perdew}\ \emph {et~al.}(1996)\citenamefont {Perdew},
  \citenamefont {Burke},\ and\ \citenamefont
  {Ernzerhof}}]{perdew_generalized_1996}%
  \BibitemOpen
  \bibfield  {author} {\bibinfo {author} {\bibfnamefont {J.~P.}\ \bibnamefont
  {Perdew}}, \bibinfo {author} {\bibfnamefont {K.}~\bibnamefont {Burke}}, \
  and\ \bibinfo {author} {\bibfnamefont {M.}~\bibnamefont {Ernzerhof}},\
  }\href@noop {} {\bibfield  {journal} {\bibinfo  {journal} {Phys. Rev. Lett.}\
  }\textbf {\bibinfo {volume} {77}},\ \bibinfo {pages} {3865} (\bibinfo {year}
  {1996})}\BibitemShut {NoStop}%
\bibitem [{\citenamefont {Sun}\ \emph {et~al.}(2015)\citenamefont {Sun},
  \citenamefont {Ruzsinszky},\ and\ \citenamefont
  {Perdew}}]{sun_strongly_2015}%
  \BibitemOpen
  \bibfield  {author} {\bibinfo {author} {\bibfnamefont {J.}~\bibnamefont
  {Sun}}, \bibinfo {author} {\bibfnamefont {A.}~\bibnamefont {Ruzsinszky}}, \
  and\ \bibinfo {author} {\bibfnamefont {J.~P.}\ \bibnamefont {Perdew}},\
  }\href {\doibase 10.1103/PhysRevLett.115.036402} {\bibfield  {journal}
  {\bibinfo  {journal} {Phys. Rev. Lett.}\ }\textbf {\bibinfo {volume} {115}},\
  \bibinfo {pages} {036402} (\bibinfo {year} {2015})}\BibitemShut {NoStop}%
\bibitem [{\citenamefont {Isaacs}\ and\ \citenamefont
  {Wolverton}(2018)}]{isaacs_performance_2018}%
  \BibitemOpen
  \bibfield  {author} {\bibinfo {author} {\bibfnamefont {E.~B.}\ \bibnamefont
  {Isaacs}}\ and\ \bibinfo {author} {\bibfnamefont {C.}~\bibnamefont
  {Wolverton}},\ }\href {\doibase 10.1103/PhysRevMaterials.2.063801} {\bibfield
   {journal} {\bibinfo  {journal} {Phys. Rev. Materials}\ }\textbf {\bibinfo
  {volume} {2}},\ \bibinfo {pages} {063801} (\bibinfo {year}
  {2018})}\BibitemShut {NoStop}%
\bibitem [{\citenamefont {Austin}\ \emph {et~al.}(2012)\citenamefont {Austin},
  \citenamefont {Zubarev},\ and\ \citenamefont {Lester}}]{austin_quantum_2012}%
  \BibitemOpen
  \bibfield  {author} {\bibinfo {author} {\bibfnamefont {B.~M.}\ \bibnamefont
  {Austin}}, \bibinfo {author} {\bibfnamefont {D.~Y.}\ \bibnamefont {Zubarev}},
  \ and\ \bibinfo {author} {\bibfnamefont {W.~A.}\ \bibnamefont {Lester}},\
  }\href {\doibase 10.1021/cr2001564} {\bibfield  {journal} {\bibinfo
  {journal} {Chem. Rev.}\ }\textbf {\bibinfo {volume} {112}},\ \bibinfo {pages}
  {263} (\bibinfo {year} {2012})}\BibitemShut {NoStop}%
\bibitem [{\citenamefont {Foulkes}\ \emph {et~al.}(2001)\citenamefont
  {Foulkes}, \citenamefont {Mitas}, \citenamefont {Needs},\ and\ \citenamefont
  {Rajagopal}}]{foulkes_quantum_2001}%
  \BibitemOpen
  \bibfield  {author} {\bibinfo {author} {\bibfnamefont {W.~M.~C.}\
  \bibnamefont {Foulkes}}, \bibinfo {author} {\bibfnamefont {L.}~\bibnamefont
  {Mitas}}, \bibinfo {author} {\bibfnamefont {R.~J.}\ \bibnamefont {Needs}}, \
  and\ \bibinfo {author} {\bibfnamefont {G.}~\bibnamefont {Rajagopal}},\ }\href
  {\doibase 10.1103/RevModPhys.73.33} {\bibfield  {journal} {\bibinfo
  {journal} {Rev. Mod. Phys.}\ }\textbf {\bibinfo {volume} {73}},\ \bibinfo
  {pages} {33} (\bibinfo {year} {2001})}\BibitemShut {NoStop}%
\bibitem [{\citenamefont {Kolorenč}\ and\ \citenamefont
  {Mitas}(2011)}]{kolorenc_applications_2011}%
  \BibitemOpen
  \bibfield  {author} {\bibinfo {author} {\bibfnamefont {J.}~\bibnamefont
  {Kolorenč}}\ and\ \bibinfo {author} {\bibfnamefont {L.}~\bibnamefont
  {Mitas}},\ }\href {\doibase 10.1088/0034-4885/74/2/026502} {\bibfield
  {journal} {\bibinfo  {journal} {Rep. Prog. Phys.}\ }\textbf {\bibinfo
  {volume} {74}},\ \bibinfo {pages} {026502} (\bibinfo {year}
  {2011})}\BibitemShut {NoStop}%
\bibitem [{\citenamefont {Toulouse}\ \emph {et~al.}(2016)\citenamefont
  {Toulouse}, \citenamefont {Assaraf},\ and\ \citenamefont
  {Umrigar}}]{toulouse_chapter_2016}%
  \BibitemOpen
  \bibfield  {author} {\bibinfo {author} {\bibfnamefont {J.}~\bibnamefont
  {Toulouse}}, \bibinfo {author} {\bibfnamefont {R.}~\bibnamefont {Assaraf}}, \
  and\ \bibinfo {author} {\bibfnamefont {C.~J.}\ \bibnamefont {Umrigar}},\ }in\
  \href {\doibase 10.1016/bs.aiq.2015.07.003} {\emph {\bibinfo {booktitle}
  {Advances in {Quantum} {Chemistry}}}},\ \bibinfo {series} {Electron
  {Correlation} in {Molecules} – ab initio {Beyond} {Gaussian} {Quantum}
  {Chemistry}}, Vol.~\bibinfo {volume} {73},\ \bibinfo {editor} {edited by\
  \bibinfo {editor} {\bibfnamefont {P.~E.}\ \bibnamefont {Hoggan}}\ and\
  \bibinfo {editor} {\bibfnamefont {T.}~\bibnamefont {Ozdogan}}}\ (\bibinfo
  {publisher} {Academic Press},\ \bibinfo {year} {2016})\ pp.\ \bibinfo {pages}
  {285--314}\BibitemShut {NoStop}%
\bibitem [{\citenamefont {Melton}\ \emph
  {et~al.}(2016{\natexlab{a}})\citenamefont {Melton}, \citenamefont {Zhu},
  \citenamefont {Guo}, \citenamefont {Ambrosetti}, \citenamefont {Pederiva},\
  and\ \citenamefont {Mitas}}]{melton_spin-orbit_2016}%
  \BibitemOpen
  \bibfield  {author} {\bibinfo {author} {\bibfnamefont {C.~A.}\ \bibnamefont
  {Melton}}, \bibinfo {author} {\bibfnamefont {M.}~\bibnamefont {Zhu}},
  \bibinfo {author} {\bibfnamefont {S.}~\bibnamefont {Guo}}, \bibinfo {author}
  {\bibfnamefont {A.}~\bibnamefont {Ambrosetti}}, \bibinfo {author}
  {\bibfnamefont {F.}~\bibnamefont {Pederiva}}, \ and\ \bibinfo {author}
  {\bibfnamefont {L.}~\bibnamefont {Mitas}},\ }\href {\doibase
  10.1103/PhysRevA.93.042502} {\bibfield  {journal} {\bibinfo  {journal} {Phys.
  Rev. A}\ }\textbf {\bibinfo {volume} {93}},\ \bibinfo {pages} {042502}
  (\bibinfo {year} {2016}{\natexlab{a}})}\BibitemShut {NoStop}%
\bibitem [{\citenamefont {Annaberdiyev}\ \emph {et~al.}(2021)\citenamefont
  {Annaberdiyev}, \citenamefont {Wang}, \citenamefont {Melton}, \citenamefont
  {Bennett},\ and\ \citenamefont {Mitas}}]{annaberdiyev_cohesion_2021}%
  \BibitemOpen
  \bibfield  {author} {\bibinfo {author} {\bibfnamefont {A.}~\bibnamefont
  {Annaberdiyev}}, \bibinfo {author} {\bibfnamefont {G.}~\bibnamefont {Wang}},
  \bibinfo {author} {\bibfnamefont {C.~A.}\ \bibnamefont {Melton}}, \bibinfo
  {author} {\bibfnamefont {M.~C.}\ \bibnamefont {Bennett}}, \ and\ \bibinfo
  {author} {\bibfnamefont {L.}~\bibnamefont {Mitas}},\ }\href {\doibase
  10.1103/PhysRevB.103.205206} {\bibfield  {journal} {\bibinfo  {journal}
  {Physical Review B}\ }\textbf {\bibinfo {volume} {103}},\ \bibinfo {pages}
  {205206} (\bibinfo {year} {2021})},\ \bibinfo {note} {publisher: American
  Physical Society}\BibitemShut {NoStop}%
\bibitem [{\citenamefont {Benali}\ \emph {et~al.}(2020)\citenamefont {Benali},
  \citenamefont {Gasperich}, \citenamefont {Jordan}, \citenamefont
  {Applencourt}, \citenamefont {Luo}, \citenamefont {Bennett}, \citenamefont
  {Krogel}, \citenamefont {Shulenburger}, \citenamefont {Kent}, \citenamefont
  {Loos}, \citenamefont {Scemama},\ and\ \citenamefont
  {Caffarel}}]{benali_toward_2020}%
  \BibitemOpen
  \bibfield  {author} {\bibinfo {author} {\bibfnamefont {A.}~\bibnamefont
  {Benali}}, \bibinfo {author} {\bibfnamefont {K.}~\bibnamefont {Gasperich}},
  \bibinfo {author} {\bibfnamefont {K.~D.}\ \bibnamefont {Jordan}}, \bibinfo
  {author} {\bibfnamefont {T.}~\bibnamefont {Applencourt}}, \bibinfo {author}
  {\bibfnamefont {Y.}~\bibnamefont {Luo}}, \bibinfo {author} {\bibfnamefont
  {M.~C.}\ \bibnamefont {Bennett}}, \bibinfo {author} {\bibfnamefont {J.~T.}\
  \bibnamefont {Krogel}}, \bibinfo {author} {\bibfnamefont {L.}~\bibnamefont
  {Shulenburger}}, \bibinfo {author} {\bibfnamefont {P.~R.~C.}\ \bibnamefont
  {Kent}}, \bibinfo {author} {\bibfnamefont {P.-F.}\ \bibnamefont {Loos}},
  \bibinfo {author} {\bibfnamefont {A.}~\bibnamefont {Scemama}}, \ and\
  \bibinfo {author} {\bibfnamefont {M.}~\bibnamefont {Caffarel}},\ }\href
  {\doibase 10.1063/5.0021036} {\bibfield  {journal} {\bibinfo  {journal} {The
  Journal of Chemical Physics}\ }\textbf {\bibinfo {volume} {153}},\ \bibinfo
  {pages} {184111} (\bibinfo {year} {2020})},\ \bibinfo {note} {publisher:
  American Institute of Physics}\BibitemShut {NoStop}%
\bibitem [{\citenamefont {Reboredo}\ and\ \citenamefont
  {Williamson}(2005)}]{reboredo_optimized_2005}%
  \BibitemOpen
  \bibfield  {author} {\bibinfo {author} {\bibfnamefont {F.~A.}\ \bibnamefont
  {Reboredo}}\ and\ \bibinfo {author} {\bibfnamefont {A.~J.}\ \bibnamefont
  {Williamson}},\ }\href {\doibase 10.1103/PhysRevB.71.121105} {\bibfield
  {journal} {\bibinfo  {journal} {Physical Review B}\ }\textbf {\bibinfo
  {volume} {71}},\ \bibinfo {pages} {121105} (\bibinfo {year}
  {2005})}\BibitemShut {NoStop}%
\bibitem [{\citenamefont {Melton}\ \emph
  {et~al.}(2016{\natexlab{b}})\citenamefont {Melton}, \citenamefont {Bennett},\
  and\ \citenamefont {Mitas}}]{melton_quantum_2016}%
  \BibitemOpen
  \bibfield  {author} {\bibinfo {author} {\bibfnamefont {C.~A.}\ \bibnamefont
  {Melton}}, \bibinfo {author} {\bibfnamefont {M.~C.}\ \bibnamefont {Bennett}},
  \ and\ \bibinfo {author} {\bibfnamefont {L.}~\bibnamefont {Mitas}},\ }\href
  {\doibase 10.1063/1.4954726} {\bibfield  {journal} {\bibinfo  {journal} {J.
  Chem. Phys.}\ }\textbf {\bibinfo {volume} {144}},\ \bibinfo {pages} {244113}
  (\bibinfo {year} {2016}{\natexlab{b}})}\BibitemShut {NoStop}%
\bibitem [{\citenamefont {Melton}\ \emph {et~al.}(2019)\citenamefont {Melton},
  \citenamefont {Bennett},\ and\ \citenamefont
  {Mitas}}]{melton_projector_2019}%
  \BibitemOpen
  \bibfield  {author} {\bibinfo {author} {\bibfnamefont {C.~A.}\ \bibnamefont
  {Melton}}, \bibinfo {author} {\bibfnamefont {M.~C.}\ \bibnamefont {Bennett}},
  \ and\ \bibinfo {author} {\bibfnamefont {L.}~\bibnamefont {Mitas}},\ }\href
  {\doibase 10.1016/j.jpcs.2017.12.033} {\bibfield  {journal} {\bibinfo
  {journal} {Journal of Physics and Chemistry of Solids}\ }\bibinfo {series}
  {Spin-{Orbit} {Coupled} {Materials}},\ \textbf {\bibinfo {volume} {128}},\
  \bibinfo {pages} {367} (\bibinfo {year} {2019})}\BibitemShut {NoStop}%
\bibitem [{\citenamefont {Giessen}\ and\ \citenamefont
  {Grant}(1965)}]{giessen_crystal_1965}%
  \BibitemOpen
  \bibfield  {author} {\bibinfo {author} {\bibfnamefont {B.~C.}\ \bibnamefont
  {Giessen}}\ and\ \bibinfo {author} {\bibfnamefont {N.~J.}\ \bibnamefont
  {Grant}},\ }\href {\doibase 10.1016/0022-5088(65)90102-5} {\bibfield
  {journal} {\bibinfo  {journal} {J. Less Common Metals}\ }\textbf {\bibinfo
  {volume} {8}},\ \bibinfo {pages} {114} (\bibinfo {year} {1965})}\BibitemShut
  {NoStop}%
\bibitem [{\citenamefont
  {Waterstrat}(1973)}]{waterstrat_vanadium-platinum_1973}%
  \BibitemOpen
  \bibfield  {author} {\bibinfo {author} {\bibfnamefont {R.~M.}\ \bibnamefont
  {Waterstrat}},\ }\href {\doibase 10.1007/BF02648698} {\bibfield  {journal}
  {\bibinfo  {journal} {Metall. Trans.}\ }\textbf {\bibinfo {volume} {4}},\
  \bibinfo {pages} {455} (\bibinfo {year} {1973})}\BibitemShut {NoStop}%
\bibitem [{\citenamefont {Wolverton}\ \emph {et~al.}(1993)\citenamefont
  {Wolverton}, \citenamefont {Ceder}, \citenamefont {de~Fontaine},\ and\
  \citenamefont {Dreysse}}]{wolverton_ab_1993}%
  \BibitemOpen
  \bibfield  {author} {\bibinfo {author} {\bibfnamefont {C.}~\bibnamefont
  {Wolverton}}, \bibinfo {author} {\bibfnamefont {G.}~\bibnamefont {Ceder}},
  \bibinfo {author} {\bibfnamefont {D.}~\bibnamefont {de~Fontaine}}, \ and\
  \bibinfo {author} {\bibfnamefont {H.}~\bibnamefont {Dreysse}},\ }\href
  {\doibase 10.1103/PhysRevB.48.726} {\bibfield  {journal} {\bibinfo  {journal}
  {Phys. Rev. B}\ }\textbf {\bibinfo {volume} {48}},\ \bibinfo {pages} {726}
  (\bibinfo {year} {1993})}\BibitemShut {NoStop}%
\bibitem [{\citenamefont {Guo}\ and\ \citenamefont
  {Kleppa}(1994)}]{guo_standard_1994}%
  \BibitemOpen
  \bibfield  {author} {\bibinfo {author} {\bibfnamefont {Q.}~\bibnamefont
  {Guo}}\ and\ \bibinfo {author} {\bibfnamefont {O.~J.}\ \bibnamefont
  {Kleppa}},\ }\href {\doibase 10.1016/0925-8388(94)90767-6} {\bibfield
  {journal} {\bibinfo  {journal} {J. Alloys Compd.}\ }\textbf {\bibinfo
  {volume} {205}},\ \bibinfo {pages} {63} (\bibinfo {year} {1994})}\BibitemShut
  {NoStop}%
\bibitem [{\citenamefont {Kim}\ \emph {et~al.}(2017)\citenamefont {Kim},
  \citenamefont {Meschel}, \citenamefont {Nash},\ and\ \citenamefont
  {Chen}}]{kim_experimental_2017}%
  \BibitemOpen
  \bibfield  {author} {\bibinfo {author} {\bibfnamefont {G.}~\bibnamefont
  {Kim}}, \bibinfo {author} {\bibfnamefont {S.~V.}\ \bibnamefont {Meschel}},
  \bibinfo {author} {\bibfnamefont {P.}~\bibnamefont {Nash}}, \ and\ \bibinfo
  {author} {\bibfnamefont {W.}~\bibnamefont {Chen}},\ }\href {\doibase
  10.1038/sdata.2017.162} {\bibfield  {journal} {\bibinfo  {journal}
  {Scientific data}\ }\textbf {\bibinfo {volume} {4}},\ \bibinfo {pages} {1}
  (\bibinfo {year} {2017})}\BibitemShut {NoStop}%
\bibitem [{\citenamefont {Nash}\ \emph {et~al.}(2020)\citenamefont {Nash},
  \citenamefont {Meschel},\ and\ \citenamefont {Gu}}]{nash_two_2020}%
  \BibitemOpen
  \bibfield  {author} {\bibinfo {author} {\bibfnamefont {P.}~\bibnamefont
  {Nash}}, \bibinfo {author} {\bibfnamefont {S.}~\bibnamefont {Meschel}}, \
  and\ \bibinfo {author} {\bibfnamefont {Q.}~\bibnamefont {Gu}},\ }\href
  {\doibase 10.1134/S003602442013018X} {\bibfield  {journal} {\bibinfo
  {journal} {Russ. J. Phys. Chem.}\ }\textbf {\bibinfo {volume} {94}},\
  \bibinfo {pages} {2624} (\bibinfo {year} {2020})}\BibitemShut {NoStop}%
\bibitem [{\citenamefont {Meschel}(2020)}]{meschel_brief_2020}%
  \BibitemOpen
  \bibfield  {author} {\bibinfo {author} {\bibfnamefont {S.~V.}\ \bibnamefont
  {Meschel}},\ }\href {\doibase 10.1016/j.calphad.2019.101714} {\bibfield
  {journal} {\bibinfo  {journal} {Calphad}\ }\textbf {\bibinfo {volume} {68}},\
  \bibinfo {pages} {101714} (\bibinfo {year} {2020})}\BibitemShut {NoStop}%
\bibitem [{\citenamefont {Funke}(1976)}]{funke_agi-type_1976}%
  \BibitemOpen
  \bibfield  {author} {\bibinfo {author} {\bibfnamefont {K.}~\bibnamefont
  {Funke}},\ }\href {\doibase 10.1016/0079-6786(76)90001-7} {\bibfield
  {journal} {\bibinfo  {journal} {Prog. Solid State Chem.}\ }\textbf {\bibinfo
  {volume} {11}},\ \bibinfo {pages} {345} (\bibinfo {year} {1976})}\BibitemShut
  {NoStop}%
\bibitem [{\citenamefont {Shan}\ \emph {et~al.}(2009)\citenamefont {Shan},
  \citenamefont {Li}, \citenamefont {Tian}, \citenamefont {Han}, \citenamefont
  {Wang}, \citenamefont {Liu}, \citenamefont {Du},\ and\ \citenamefont
  {Yang}}]{shan_description_2009}%
  \BibitemOpen
  \bibfield  {author} {\bibinfo {author} {\bibfnamefont {Y.}~\bibnamefont
  {Shan}}, \bibinfo {author} {\bibfnamefont {G.}~\bibnamefont {Li}}, \bibinfo
  {author} {\bibfnamefont {G.}~\bibnamefont {Tian}}, \bibinfo {author}
  {\bibfnamefont {J.}~\bibnamefont {Han}}, \bibinfo {author} {\bibfnamefont
  {C.}~\bibnamefont {Wang}}, \bibinfo {author} {\bibfnamefont {S.}~\bibnamefont
  {Liu}}, \bibinfo {author} {\bibfnamefont {H.}~\bibnamefont {Du}}, \ and\
  \bibinfo {author} {\bibfnamefont {Y.}~\bibnamefont {Yang}},\ }\href {\doibase
  10.1016/j.jallcom.2008.10.026} {\bibfield  {journal} {\bibinfo  {journal} {J.
  Alloys Compd.}\ }\textbf {\bibinfo {volume} {477}},\ \bibinfo {pages} {403}
  (\bibinfo {year} {2009})}\BibitemShut {NoStop}%
\bibitem [{\citenamefont {Grundmann}\ \emph {et~al.}(2013)\citenamefont
  {Grundmann}, \citenamefont {Schein}, \citenamefont {Lorenz}, \citenamefont
  {Böntgen}, \citenamefont {Lenzner},\ and\ \citenamefont
  {Wenckstern}}]{grundmann_cuprous_2013}%
  \BibitemOpen
  \bibfield  {author} {\bibinfo {author} {\bibfnamefont {M.}~\bibnamefont
  {Grundmann}}, \bibinfo {author} {\bibfnamefont {F.-L.}\ \bibnamefont
  {Schein}}, \bibinfo {author} {\bibfnamefont {M.}~\bibnamefont {Lorenz}},
  \bibinfo {author} {\bibfnamefont {T.}~\bibnamefont {Böntgen}}, \bibinfo
  {author} {\bibfnamefont {J.}~\bibnamefont {Lenzner}}, \ and\ \bibinfo
  {author} {\bibfnamefont {H.~v.}\ \bibnamefont {Wenckstern}},\ }\href
  {\doibase 10.1002/pssa.201329349} {\bibfield  {journal} {\bibinfo  {journal}
  {Phys. Status Solidi A}\ }\textbf {\bibinfo {volume} {210}},\ \bibinfo
  {pages} {1671} (\bibinfo {year} {2013})}\BibitemShut {NoStop}%
\bibitem [{\citenamefont {Miyake}\ \emph {et~al.}(1952)\citenamefont {Miyake},
  \citenamefont {Hoshino},\ and\ \citenamefont {Takenaka}}]{miyake_phase_1952}%
  \BibitemOpen
  \bibfield  {author} {\bibinfo {author} {\bibfnamefont {S.}~\bibnamefont
  {Miyake}}, \bibinfo {author} {\bibfnamefont {S.}~\bibnamefont {Hoshino}}, \
  and\ \bibinfo {author} {\bibfnamefont {T.}~\bibnamefont {Takenaka}},\ }\href
  {\doibase 10.1143/JPSJ.7.19} {\bibfield  {journal} {\bibinfo  {journal} {J.
  Phys. Soc. Jpn.}\ }\textbf {\bibinfo {volume} {7}},\ \bibinfo {pages} {19}
  (\bibinfo {year} {1952})}\BibitemShut {NoStop}%
\bibitem [{\citenamefont {Rapoport}\ and\ \citenamefont
  {Pistorius}(1968)}]{rapoport_phase_1968}%
  \BibitemOpen
  \bibfield  {author} {\bibinfo {author} {\bibfnamefont {E.}~\bibnamefont
  {Rapoport}}\ and\ \bibinfo {author} {\bibfnamefont {C.~W. F.~T.}\
  \bibnamefont {Pistorius}},\ }\href {\doibase 10.1103/PhysRev.172.838}
  {\bibfield  {journal} {\bibinfo  {journal} {Phys. Rev.}\ }\textbf {\bibinfo
  {volume} {172}},\ \bibinfo {pages} {838} (\bibinfo {year}
  {1968})}\BibitemShut {NoStop}%
\bibitem [{\citenamefont {of~Science}\ and\ \citenamefont
  {University}()}]{russian_database}%
  \BibitemOpen
  \bibfield  {author} {\bibinfo {author} {\bibfnamefont {R.~A.}\ \bibnamefont
  {of~Science}}\ and\ \bibinfo {author} {\bibfnamefont {M.~S.}\ \bibnamefont
  {University}},\ }\href@noop {} {\enquote {\bibinfo {title} {Thermodynamic
  properties of individual substances},}\ }\bibinfo {howpublished}
  {\url{http://www.chem.msu.su/rus/tsiv/Cu/print-CuI\_c.html}},\ \bibinfo
  {note} {accessed: 2020-04-01}\BibitemShut {NoStop}%
\bibitem [{\citenamefont {Belov}\ \emph {et~al.}(2018)\citenamefont {Belov},
  \citenamefont {Dyachkov}, \citenamefont {Levashov}, \citenamefont
  {Lomonosov}, \citenamefont {Minakov}, \citenamefont {Morozov}, \citenamefont
  {Sineva},\ and\ \citenamefont {Smirnov}}]{belov2018ivtanthermo}%
  \BibitemOpen
  \bibfield  {author} {\bibinfo {author} {\bibfnamefont {G.}~\bibnamefont
  {Belov}}, \bibinfo {author} {\bibfnamefont {S.}~\bibnamefont {Dyachkov}},
  \bibinfo {author} {\bibfnamefont {P.}~\bibnamefont {Levashov}}, \bibinfo
  {author} {\bibfnamefont {I.}~\bibnamefont {Lomonosov}}, \bibinfo {author}
  {\bibfnamefont {D.}~\bibnamefont {Minakov}}, \bibinfo {author} {\bibfnamefont
  {I.}~\bibnamefont {Morozov}}, \bibinfo {author} {\bibfnamefont
  {M.}~\bibnamefont {Sineva}}, \ and\ \bibinfo {author} {\bibfnamefont
  {V.}~\bibnamefont {Smirnov}},\ }in\ \href@noop {} {\emph {\bibinfo
  {booktitle} {Journal of Physics: Conference Series}}},\ Vol.\ \bibinfo
  {volume} {946}\ (\bibinfo {organization} {IOP Publishing},\ \bibinfo {year}
  {2018})\ p.\ \bibinfo {pages} {012120}\BibitemShut {NoStop}%
\bibitem [{\citenamefont {Cartwright}\ and\ \citenamefont
  {A. Woolf}(1976)}]{cartwright_heats_1976}%
  \BibitemOpen
  \bibfield  {author} {\bibinfo {author} {\bibfnamefont {M.}~\bibnamefont
  {Cartwright}}\ and\ \bibinfo {author} {\bibfnamefont {A.}~\bibnamefont
  {A. Woolf}},\ }\href {\doibase 10.1039/DT9760000829} {\bibfield  {journal}
  {\bibinfo  {journal} {Dalton Trans.}\ }\textbf {\bibinfo {volume} {9}},\
  \bibinfo {pages} {829} (\bibinfo {year} {1976})}\BibitemShut {NoStop}%
\bibitem [{\citenamefont {Ishikawa}\ \emph {et~al.}(1934)\citenamefont
  {Ishikawa}, \citenamefont {Yamazaki},\ and\ \citenamefont
  {Murooka}}]{ishikawa_studies_1934}%
  \BibitemOpen
  \bibfield  {author} {\bibinfo {author} {\bibfnamefont {F.}~\bibnamefont
  {Ishikawa}}, \bibinfo {author} {\bibfnamefont {S.}~\bibnamefont {Yamazaki}},
  \ and\ \bibinfo {author} {\bibfnamefont {T.}~\bibnamefont {Murooka}},\
  }\href@noop {} {\bibfield  {journal} {\bibinfo  {journal} {Sci. Repts. Tohoku
  Imp. Univ.}\ }\textbf {\bibinfo {volume} {1}},\ \bibinfo {pages} {115}
  (\bibinfo {year} {1934})}\BibitemShut {NoStop}%
\bibitem [{\citenamefont {Dinsdale}(1991)}]{dinsdale_sgte_1991}%
  \BibitemOpen
  \bibfield  {author} {\bibinfo {author} {\bibfnamefont {A.~T.}\ \bibnamefont
  {Dinsdale}},\ }\href {\doibase 10.1016/0364-5916(91)90030-N} {\bibfield
  {journal} {\bibinfo  {journal} {Calphad}\ }\textbf {\bibinfo {volume} {15}},\
  \bibinfo {pages} {317} (\bibinfo {year} {1991})}\BibitemShut {NoStop}%
\bibitem [{\citenamefont {Yamamoto}\ \emph {et~al.}(1987)\citenamefont
  {Yamamoto}, \citenamefont {Seki}, \citenamefont {Mori},\ and\ \citenamefont
  {Inokuchi}}]{yamamoto_ultraviolet_1987}%
  \BibitemOpen
  \bibfield  {author} {\bibinfo {author} {\bibfnamefont {H.}~\bibnamefont
  {Yamamoto}}, \bibinfo {author} {\bibfnamefont {K.}~\bibnamefont {Seki}},
  \bibinfo {author} {\bibfnamefont {T.}~\bibnamefont {Mori}}, \ and\ \bibinfo
  {author} {\bibfnamefont {H.}~\bibnamefont {Inokuchi}},\ }\href {\doibase
  10.1063/1.452177} {\bibfield  {journal} {\bibinfo  {journal} {J. Chem.
  Phys.}\ }\textbf {\bibinfo {volume} {86}},\ \bibinfo {pages} {1775} (\bibinfo
  {year} {1987})}\BibitemShut {NoStop}%
\bibitem [{\citenamefont {van Bolhuis}\ \emph {et~al.}(1967)\citenamefont {van
  Bolhuis}, \citenamefont {Koster},\ and\ \citenamefont
  {Migchelsen}}]{van_bolhuis_refinement_1967}%
  \BibitemOpen
  \bibfield  {author} {\bibinfo {author} {\bibfnamefont {F.}~\bibnamefont {van
  Bolhuis}}, \bibinfo {author} {\bibfnamefont {P.~B.}\ \bibnamefont {Koster}},
  \ and\ \bibinfo {author} {\bibfnamefont {T.}~\bibnamefont {Migchelsen}},\
  }\href {\doibase 10.1107/S0365110X6700218X} {\bibfield  {journal} {\bibinfo
  {journal} {Acta Cryst.}\ }\textbf {\bibinfo {volume} {23}},\ \bibinfo {pages}
  {90} (\bibinfo {year} {1967})}\BibitemShut {NoStop}%
\bibitem [{\citenamefont {Stepanov}(2005)}]{stepanov_iodine_2005}%
  \BibitemOpen
  \bibfield  {author} {\bibinfo {author} {\bibfnamefont {I.~A.}\ \bibnamefont
  {Stepanov}},\ }\href {\doibase 10.1134/1.2087735} {\bibfield  {journal}
  {\bibinfo  {journal} {Phys. Solid State}\ }\textbf {\bibinfo {volume} {47}},\
  \bibinfo {pages} {1852} (\bibinfo {year} {2005})}\BibitemShut {NoStop}%
\bibitem [{\citenamefont {Rumble}(2020)}]{crc_handbook}%
  \BibitemOpen
  \bibinfo {editor} {\bibfnamefont {J.~R.}\ \bibnamefont {Rumble}},\ ed.,\
  \href@noop {} {\emph {\bibinfo {title} {{CRC Handbook of Chemistry and
  Physics}}}}\ (\bibinfo  {publisher} {CRC Press/Taylor \& Francis, Boca Raton,
  FL.},\ \bibinfo {year} {2020})\ \bibinfo {note} {101st Edition (Internet
  Version)}\BibitemShut {NoStop}%
\bibitem [{\citenamefont {Edwards}\ \emph {et~al.}(1951)\citenamefont
  {Edwards}, \citenamefont {Speiser},\ and\ \citenamefont
  {Johnston}}]{edwards_high_1951}%
  \BibitemOpen
  \bibfield  {author} {\bibinfo {author} {\bibfnamefont {J.~W.}\ \bibnamefont
  {Edwards}}, \bibinfo {author} {\bibfnamefont {R.}~\bibnamefont {Speiser}}, \
  and\ \bibinfo {author} {\bibfnamefont {H.~L.}\ \bibnamefont {Johnston}},\
  }\href {\doibase 10.1063/1.1699977} {\bibfield  {journal} {\bibinfo
  {journal} {J. Appl. Phys.}\ }\textbf {\bibinfo {volume} {22}},\ \bibinfo
  {pages} {424} (\bibinfo {year} {1951})}\BibitemShut {NoStop}%
\bibitem [{\citenamefont {Davey}(1925)}]{davey_precision_1925}%
  \BibitemOpen
  \bibfield  {author} {\bibinfo {author} {\bibfnamefont {W.~P.}\ \bibnamefont
  {Davey}},\ }\href {\doibase 10.1103/PhysRev.25.753} {\bibfield  {journal}
  {\bibinfo  {journal} {Phys. Rev.}\ }\textbf {\bibinfo {volume} {25}},\
  \bibinfo {pages} {753} (\bibinfo {year} {1925})}\BibitemShut {NoStop}%
\bibitem [{\citenamefont {Krogel}\ \emph {et~al.}(2016)\citenamefont {Krogel},
  \citenamefont {Santana},\ and\ \citenamefont
  {Reboredo}}]{krogel_pseudopotentials_2016}%
  \BibitemOpen
  \bibfield  {author} {\bibinfo {author} {\bibfnamefont {J.~T.}\ \bibnamefont
  {Krogel}}, \bibinfo {author} {\bibfnamefont {J.~A.}\ \bibnamefont {Santana}},
  \ and\ \bibinfo {author} {\bibfnamefont {F.~A.}\ \bibnamefont {Reboredo}},\
  }\href {\doibase 10.1103/PhysRevB.93.075143} {\bibfield  {journal} {\bibinfo
  {journal} {Phys. Rev. B}\ }\textbf {\bibinfo {volume} {93}},\ \bibinfo
  {pages} {075143} (\bibinfo {year} {2016})}\BibitemShut {NoStop}%
\bibitem [{\citenamefont {Rappe}\ \emph {et~al.}(1990)\citenamefont {Rappe},
  \citenamefont {Rabe}, \citenamefont {Kaxiras},\ and\ \citenamefont
  {Joannopoulos}}]{rappe_optimized_1990}%
  \BibitemOpen
  \bibfield  {author} {\bibinfo {author} {\bibfnamefont {A.~M.}\ \bibnamefont
  {Rappe}}, \bibinfo {author} {\bibfnamefont {K.~M.}\ \bibnamefont {Rabe}},
  \bibinfo {author} {\bibfnamefont {E.}~\bibnamefont {Kaxiras}}, \ and\
  \bibinfo {author} {\bibfnamefont {J.~D.}\ \bibnamefont {Joannopoulos}},\
  }\href {\doibase 10.1103/PhysRevB.41.1227} {\bibfield  {journal} {\bibinfo
  {journal} {Phys. Rev. B}\ }\textbf {\bibinfo {volume} {41}},\ \bibinfo
  {pages} {1227} (\bibinfo {year} {1990})}\BibitemShut {NoStop}%
\bibitem [{\citenamefont {Troullier}\ and\ \citenamefont
  {Martins}(1991)}]{troullier_efficient_1991}%
  \BibitemOpen
  \bibfield  {author} {\bibinfo {author} {\bibfnamefont {N.}~\bibnamefont
  {Troullier}}\ and\ \bibinfo {author} {\bibfnamefont {J.~L.}\ \bibnamefont
  {Martins}},\ }\href {\doibase 10.1103/PhysRevB.43.1993} {\bibfield  {journal}
  {\bibinfo  {journal} {Phys. Rev. B}\ }\textbf {\bibinfo {volume} {43}},\
  \bibinfo {pages} {1993} (\bibinfo {year} {1991})}\BibitemShut {NoStop}%
\bibitem [{\citenamefont {Burkatzki}\ \emph {et~al.}(2007)\citenamefont
  {Burkatzki}, \citenamefont {Filippi},\ and\ \citenamefont
  {Dolg}}]{burkatzki_energy-consistent_2007}%
  \BibitemOpen
  \bibfield  {author} {\bibinfo {author} {\bibfnamefont {M.}~\bibnamefont
  {Burkatzki}}, \bibinfo {author} {\bibfnamefont {C.}~\bibnamefont {Filippi}},
  \ and\ \bibinfo {author} {\bibfnamefont {M.}~\bibnamefont {Dolg}},\ }\href
  {\doibase 10.1063/1.2741534} {\bibfield  {journal} {\bibinfo  {journal} {J.
  Chem. Phys.}\ }\textbf {\bibinfo {volume} {126}},\ \bibinfo {pages} {234105}
  (\bibinfo {year} {2007})}\BibitemShut {NoStop}%
\bibitem [{\citenamefont {Stoll}\ \emph {et~al.}(2002)\citenamefont {Stoll},
  \citenamefont {Metz},\ and\ \citenamefont {Dolg}}]{STU}%
  \BibitemOpen
  \bibfield  {author} {\bibinfo {author} {\bibfnamefont {H.}~\bibnamefont
  {Stoll}}, \bibinfo {author} {\bibfnamefont {B.}~\bibnamefont {Metz}}, \ and\
  \bibinfo {author} {\bibfnamefont {M.}~\bibnamefont {Dolg}},\ }\href {\doibase
  https://doi.org/10.1002/jcc.10037} {\bibfield  {journal} {\bibinfo  {journal}
  {Journal of Computational Chemistry}\ }\textbf {\bibinfo {volume} {23}},\
  \bibinfo {pages} {767} (\bibinfo {year} {2002})},\ \Eprint
  {http://arxiv.org/abs/https://onlinelibrary.wiley.com/doi/pdf/10.1002/jcc.10037}
  {https://onlinelibrary.wiley.com/doi/pdf/10.1002/jcc.10037} \BibitemShut
  {NoStop}%
\bibitem [{\citenamefont {Figgen}\ \emph {et~al.}(2009)\citenamefont {Figgen},
  \citenamefont {Peterson}, \citenamefont {Dolg},\ and\ \citenamefont
  {Stoll}}]{STU_Pt}%
  \BibitemOpen
  \bibfield  {author} {\bibinfo {author} {\bibfnamefont {D.}~\bibnamefont
  {Figgen}}, \bibinfo {author} {\bibfnamefont {K.~A.}\ \bibnamefont
  {Peterson}}, \bibinfo {author} {\bibfnamefont {M.}~\bibnamefont {Dolg}}, \
  and\ \bibinfo {author} {\bibfnamefont {H.}~\bibnamefont {Stoll}},\ }\href
  {\doibase 10.1063/1.3119665} {\bibfield  {journal} {\bibinfo  {journal} {The
  Journal of Chemical Physics}\ }\textbf {\bibinfo {volume} {130}},\ \bibinfo
  {pages} {164108} (\bibinfo {year} {2009})},\ \Eprint
  {http://arxiv.org/abs/https://doi.org/10.1063/1.3119665}
  {https://doi.org/10.1063/1.3119665} \BibitemShut {NoStop}%
\bibitem [{\citenamefont {Giannozzi}\ \emph {et~al.}(2017)\citenamefont
  {Giannozzi}, \citenamefont {Andreussi}, \citenamefont {Brumme}, \citenamefont
  {Bunau}, \citenamefont {Nardelli}, \citenamefont {Calandra}, \citenamefont
  {Car}, \citenamefont {Cavazzoni}, \citenamefont {Ceresoli}, \citenamefont
  {Cococcioni}, \citenamefont {Colonna}, \citenamefont {Carnimeo},
  \citenamefont {Corso}, \citenamefont {Gironcoli}, \citenamefont {Delugas},
  \citenamefont {DiStasio}, \citenamefont {Ferretti}, \citenamefont {Floris},
  \citenamefont {Fratesi}, \citenamefont {Fugallo}, \citenamefont {Gebauer},
  \citenamefont {Gerstmann}, \citenamefont {Giustino}, \citenamefont {Gorni},
  \citenamefont {Jia}, \citenamefont {Kawamura}, \citenamefont {Ko},
  \citenamefont {Kokalj}, \citenamefont {Küçükbenli}, \citenamefont
  {Lazzeri}, \citenamefont {Marsili}, \citenamefont {Marzari}, \citenamefont
  {Mauri}, \citenamefont {Nguyen}, \citenamefont {Nguyen}, \citenamefont
  {Otero-de-la Roza}, \citenamefont {Paulatto}, \citenamefont {Poncé},
  \citenamefont {Rocca}, \citenamefont {Sabatini}, \citenamefont {Santra},
  \citenamefont {Schlipf}, \citenamefont {Seitsonen}, \citenamefont {Smogunov},
  \citenamefont {Timrov}, \citenamefont {Thonhauser}, \citenamefont {Umari},
  \citenamefont {Vast}, \citenamefont {Wu},\ and\ \citenamefont
  {Baroni}}]{giannozzi_advanced_2017}%
  \BibitemOpen
  \bibfield  {author} {\bibinfo {author} {\bibfnamefont {P.}~\bibnamefont
  {Giannozzi}}, \bibinfo {author} {\bibfnamefont {O.}~\bibnamefont
  {Andreussi}}, \bibinfo {author} {\bibfnamefont {T.}~\bibnamefont {Brumme}},
  \bibinfo {author} {\bibfnamefont {O.}~\bibnamefont {Bunau}}, \bibinfo
  {author} {\bibfnamefont {M.~B.}\ \bibnamefont {Nardelli}}, \bibinfo {author}
  {\bibfnamefont {M.}~\bibnamefont {Calandra}}, \bibinfo {author}
  {\bibfnamefont {R.}~\bibnamefont {Car}}, \bibinfo {author} {\bibfnamefont
  {C.}~\bibnamefont {Cavazzoni}}, \bibinfo {author} {\bibfnamefont
  {D.}~\bibnamefont {Ceresoli}}, \bibinfo {author} {\bibfnamefont
  {M.}~\bibnamefont {Cococcioni}}, \bibinfo {author} {\bibfnamefont
  {N.}~\bibnamefont {Colonna}}, \bibinfo {author} {\bibfnamefont
  {I.}~\bibnamefont {Carnimeo}}, \bibinfo {author} {\bibfnamefont {A.~D.}\
  \bibnamefont {Corso}}, \bibinfo {author} {\bibfnamefont {S.~d.}\ \bibnamefont
  {Gironcoli}}, \bibinfo {author} {\bibfnamefont {P.}~\bibnamefont {Delugas}},
  \bibinfo {author} {\bibfnamefont {R.~A.}\ \bibnamefont {DiStasio}}, \bibinfo
  {author} {\bibfnamefont {A.}~\bibnamefont {Ferretti}}, \bibinfo {author}
  {\bibfnamefont {A.}~\bibnamefont {Floris}}, \bibinfo {author} {\bibfnamefont
  {G.}~\bibnamefont {Fratesi}}, \bibinfo {author} {\bibfnamefont
  {G.}~\bibnamefont {Fugallo}}, \bibinfo {author} {\bibfnamefont
  {R.}~\bibnamefont {Gebauer}}, \bibinfo {author} {\bibfnamefont
  {U.}~\bibnamefont {Gerstmann}}, \bibinfo {author} {\bibfnamefont
  {F.}~\bibnamefont {Giustino}}, \bibinfo {author} {\bibfnamefont
  {T.}~\bibnamefont {Gorni}}, \bibinfo {author} {\bibfnamefont
  {J.}~\bibnamefont {Jia}}, \bibinfo {author} {\bibfnamefont {M.}~\bibnamefont
  {Kawamura}}, \bibinfo {author} {\bibfnamefont {H.-Y.}\ \bibnamefont {Ko}},
  \bibinfo {author} {\bibfnamefont {A.}~\bibnamefont {Kokalj}}, \bibinfo
  {author} {\bibfnamefont {E.}~\bibnamefont {Küçükbenli}}, \bibinfo {author}
  {\bibfnamefont {M.}~\bibnamefont {Lazzeri}}, \bibinfo {author} {\bibfnamefont
  {M.}~\bibnamefont {Marsili}}, \bibinfo {author} {\bibfnamefont
  {N.}~\bibnamefont {Marzari}}, \bibinfo {author} {\bibfnamefont
  {F.}~\bibnamefont {Mauri}}, \bibinfo {author} {\bibfnamefont {N.~L.}\
  \bibnamefont {Nguyen}}, \bibinfo {author} {\bibfnamefont {H.-V.}\
  \bibnamefont {Nguyen}}, \bibinfo {author} {\bibfnamefont {A.}~\bibnamefont
  {Otero-de-la Roza}}, \bibinfo {author} {\bibfnamefont {L.}~\bibnamefont
  {Paulatto}}, \bibinfo {author} {\bibfnamefont {S.}~\bibnamefont {Poncé}},
  \bibinfo {author} {\bibfnamefont {D.}~\bibnamefont {Rocca}}, \bibinfo
  {author} {\bibfnamefont {R.}~\bibnamefont {Sabatini}}, \bibinfo {author}
  {\bibfnamefont {B.}~\bibnamefont {Santra}}, \bibinfo {author} {\bibfnamefont
  {M.}~\bibnamefont {Schlipf}}, \bibinfo {author} {\bibfnamefont {A.~P.}\
  \bibnamefont {Seitsonen}}, \bibinfo {author} {\bibfnamefont {A.}~\bibnamefont
  {Smogunov}}, \bibinfo {author} {\bibfnamefont {I.}~\bibnamefont {Timrov}},
  \bibinfo {author} {\bibfnamefont {T.}~\bibnamefont {Thonhauser}}, \bibinfo
  {author} {\bibfnamefont {P.}~\bibnamefont {Umari}}, \bibinfo {author}
  {\bibfnamefont {N.}~\bibnamefont {Vast}}, \bibinfo {author} {\bibfnamefont
  {X.}~\bibnamefont {Wu}}, \ and\ \bibinfo {author} {\bibfnamefont
  {S.}~\bibnamefont {Baroni}},\ }\href {\doibase 10.1088/1361-648X/aa8f79}
  {\bibfield  {journal} {\bibinfo  {journal} {J. Phys.: Condens. Matter}\
  }\textbf {\bibinfo {volume} {29}},\ \bibinfo {pages} {465901} (\bibinfo
  {year} {2017})}\BibitemShut {NoStop}%
\bibitem [{\citenamefont {Kim}\ \emph {et~al.}(2018)\citenamefont {Kim},
  \citenamefont {Baczewski}, \citenamefont {Beaudet}, \citenamefont {Benali},
  \citenamefont {Bennett}, \citenamefont {Berrill}, \citenamefont {Blunt},
  \citenamefont {Borda}, \citenamefont {Casula}, \citenamefont {Ceperley},
  \citenamefont {Chiesa}, \citenamefont {Clark}, \citenamefont {Clay},
  \citenamefont {Delaney}, \citenamefont {Dewing}, \citenamefont {Esler},
  \citenamefont {Hao}, \citenamefont {Heinonen}, \citenamefont {Kent},
  \citenamefont {Krogel}, \citenamefont {Kylanpaa}, \citenamefont {Li},
  \citenamefont {Lopez}, \citenamefont {Luo}, \citenamefont {Malone},
  \citenamefont {Martin}, \citenamefont {Mathuriya}, \citenamefont {McMinis},
  \citenamefont {Melton}, \citenamefont {Mitas}, \citenamefont {Morales},
  \citenamefont {Neuscamman}, \citenamefont {Parker}, \citenamefont {Flores},
  \citenamefont {Romero}, \citenamefont {Rubenstein}, \citenamefont {Shea},
  \citenamefont {Shin}, \citenamefont {Shulenburger}, \citenamefont {Tillack},
  \citenamefont {Townsend}, \citenamefont {Tubman}, \citenamefont {van~der
  Goetz}, \citenamefont {Vincent}, \citenamefont {Yang}, \citenamefont {Yang},
  \citenamefont {Zhang},\ and\ \citenamefont {Zhao}}]{QMCPACK}%
  \BibitemOpen
  \bibfield  {author} {\bibinfo {author} {\bibfnamefont {J.}~\bibnamefont
  {Kim}}, \bibinfo {author} {\bibfnamefont {A.}~\bibnamefont {Baczewski}},
  \bibinfo {author} {\bibfnamefont {T.}~\bibnamefont {Beaudet}}, \bibinfo
  {author} {\bibfnamefont {A.}~\bibnamefont {Benali}}, \bibinfo {author}
  {\bibfnamefont {C.}~\bibnamefont {Bennett}}, \bibinfo {author} {\bibfnamefont
  {M.}~\bibnamefont {Berrill}}, \bibinfo {author} {\bibfnamefont
  {N.}~\bibnamefont {Blunt}}, \bibinfo {author} {\bibfnamefont {E.~J.~L.}\
  \bibnamefont {Borda}}, \bibinfo {author} {\bibfnamefont {M.}~\bibnamefont
  {Casula}}, \bibinfo {author} {\bibfnamefont {D.}~\bibnamefont {Ceperley}},
  \bibinfo {author} {\bibfnamefont {S.}~\bibnamefont {Chiesa}}, \bibinfo
  {author} {\bibfnamefont {B.~K.}\ \bibnamefont {Clark}}, \bibinfo {author}
  {\bibfnamefont {R.}~\bibnamefont {Clay}}, \bibinfo {author} {\bibfnamefont
  {K.}~\bibnamefont {Delaney}}, \bibinfo {author} {\bibfnamefont
  {M.}~\bibnamefont {Dewing}}, \bibinfo {author} {\bibfnamefont
  {K.}~\bibnamefont {Esler}}, \bibinfo {author} {\bibfnamefont
  {H.}~\bibnamefont {Hao}}, \bibinfo {author} {\bibfnamefont {O.}~\bibnamefont
  {Heinonen}}, \bibinfo {author} {\bibfnamefont {P.~R.~C.}\ \bibnamefont
  {Kent}}, \bibinfo {author} {\bibfnamefont {J.~T.}\ \bibnamefont {Krogel}},
  \bibinfo {author} {\bibfnamefont {I.}~\bibnamefont {Kylanpaa}}, \bibinfo
  {author} {\bibfnamefont {Y.~W.}\ \bibnamefont {Li}}, \bibinfo {author}
  {\bibfnamefont {M.~G.}\ \bibnamefont {Lopez}}, \bibinfo {author}
  {\bibfnamefont {Y.}~\bibnamefont {Luo}}, \bibinfo {author} {\bibfnamefont
  {F.}~\bibnamefont {Malone}}, \bibinfo {author} {\bibfnamefont
  {R.}~\bibnamefont {Martin}}, \bibinfo {author} {\bibfnamefont
  {A.}~\bibnamefont {Mathuriya}}, \bibinfo {author} {\bibfnamefont
  {J.}~\bibnamefont {McMinis}}, \bibinfo {author} {\bibfnamefont
  {C.}~\bibnamefont {Melton}}, \bibinfo {author} {\bibfnamefont
  {L.}~\bibnamefont {Mitas}}, \bibinfo {author} {\bibfnamefont {M.~A.}\
  \bibnamefont {Morales}}, \bibinfo {author} {\bibfnamefont {E.}~\bibnamefont
  {Neuscamman}}, \bibinfo {author} {\bibfnamefont {W.}~\bibnamefont {Parker}},
  \bibinfo {author} {\bibfnamefont {S.}~\bibnamefont {Flores}}, \bibinfo
  {author} {\bibfnamefont {N.~A.}\ \bibnamefont {Romero}}, \bibinfo {author}
  {\bibfnamefont {B.}~\bibnamefont {Rubenstein}}, \bibinfo {author}
  {\bibfnamefont {J.}~\bibnamefont {Shea}}, \bibinfo {author} {\bibfnamefont
  {H.}~\bibnamefont {Shin}}, \bibinfo {author} {\bibfnamefont {L.}~\bibnamefont
  {Shulenburger}}, \bibinfo {author} {\bibfnamefont {A.}~\bibnamefont
  {Tillack}}, \bibinfo {author} {\bibfnamefont {J.}~\bibnamefont {Townsend}},
  \bibinfo {author} {\bibfnamefont {N.}~\bibnamefont {Tubman}}, \bibinfo
  {author} {\bibfnamefont {B.}~\bibnamefont {van~der Goetz}}, \bibinfo {author}
  {\bibfnamefont {J.}~\bibnamefont {Vincent}}, \bibinfo {author} {\bibfnamefont
  {D.~C.}\ \bibnamefont {Yang}}, \bibinfo {author} {\bibfnamefont
  {Y.}~\bibnamefont {Yang}}, \bibinfo {author} {\bibfnamefont {S.}~\bibnamefont
  {Zhang}}, \ and\ \bibinfo {author} {\bibfnamefont {L.}~\bibnamefont {Zhao}},\
  }\href {\doibase 10.1088/1361-648X/aab9c3} {\bibfield  {journal} {\bibinfo
  {journal} {J. Phys.: Condens. Matter}\ }\textbf {\bibinfo {volume} {30}},\
  \bibinfo {pages} {195901} (\bibinfo {year} {2018})}\BibitemShut {NoStop}%
\bibitem [{\citenamefont {Jastrow}(1955)}]{jastrow_many-body_1955}%
  \BibitemOpen
  \bibfield  {author} {\bibinfo {author} {\bibfnamefont {R.}~\bibnamefont
  {Jastrow}},\ }\href {\doibase 10.1103/PhysRev.98.1479} {\bibfield  {journal}
  {\bibinfo  {journal} {Phys. Rev.}\ }\textbf {\bibinfo {volume} {98}},\
  \bibinfo {pages} {1479} (\bibinfo {year} {1955})}\BibitemShut {NoStop}%
\bibitem [{\citenamefont {Drummond}\ \emph {et~al.}(2004)\citenamefont
  {Drummond}, \citenamefont {Towler},\ and\ \citenamefont
  {Needs}}]{drummond_jastrow_2004}%
  \BibitemOpen
  \bibfield  {author} {\bibinfo {author} {\bibfnamefont {N.~D.}\ \bibnamefont
  {Drummond}}, \bibinfo {author} {\bibfnamefont {M.~D.}\ \bibnamefont
  {Towler}}, \ and\ \bibinfo {author} {\bibfnamefont {R.~J.}\ \bibnamefont
  {Needs}},\ }\href {\doibase 10.1103/PhysRevB.70.235119} {\bibfield  {journal}
  {\bibinfo  {journal} {Phys. Rev. B}\ }\textbf {\bibinfo {volume} {70}},\
  \bibinfo {pages} {235119} (\bibinfo {year} {2004})}\BibitemShut {NoStop}%
\bibitem [{\citenamefont {Williamson}\ \emph {et~al.}(1997)\citenamefont
  {Williamson}, \citenamefont {Rajagopal}, \citenamefont {Needs}, \citenamefont
  {Fraser}, \citenamefont {Foulkes}, \citenamefont {Wang},\ and\ \citenamefont
  {Chou}}]{williamson_elimination_1997}%
  \BibitemOpen
  \bibfield  {author} {\bibinfo {author} {\bibfnamefont {A.~J.}\ \bibnamefont
  {Williamson}}, \bibinfo {author} {\bibfnamefont {G.}~\bibnamefont
  {Rajagopal}}, \bibinfo {author} {\bibfnamefont {R.~J.}\ \bibnamefont
  {Needs}}, \bibinfo {author} {\bibfnamefont {L.~M.}\ \bibnamefont {Fraser}},
  \bibinfo {author} {\bibfnamefont {W.~M.~C.}\ \bibnamefont {Foulkes}},
  \bibinfo {author} {\bibfnamefont {Y.}~\bibnamefont {Wang}}, \ and\ \bibinfo
  {author} {\bibfnamefont {M.-Y.}\ \bibnamefont {Chou}},\ }\href {\doibase
  10.1103/PhysRevB.55.R4851} {\bibfield  {journal} {\bibinfo  {journal} {Phys.
  Rev. B}\ }\textbf {\bibinfo {volume} {55}},\ \bibinfo {pages} {R4851}
  (\bibinfo {year} {1997})}\BibitemShut {NoStop}%
\bibitem [{\citenamefont {Kent}\ \emph {et~al.}(1999)\citenamefont {Kent},
  \citenamefont {Hood}, \citenamefont {Williamson}, \citenamefont {Needs},
  \citenamefont {Foulkes},\ and\ \citenamefont
  {Rajagopal}}]{kent_finite-size_1999}%
  \BibitemOpen
  \bibfield  {author} {\bibinfo {author} {\bibfnamefont {P.~R.~C.}\
  \bibnamefont {Kent}}, \bibinfo {author} {\bibfnamefont {R.~Q.}\ \bibnamefont
  {Hood}}, \bibinfo {author} {\bibfnamefont {A.~J.}\ \bibnamefont
  {Williamson}}, \bibinfo {author} {\bibfnamefont {R.~J.}\ \bibnamefont
  {Needs}}, \bibinfo {author} {\bibfnamefont {W.~M.~C.}\ \bibnamefont
  {Foulkes}}, \ and\ \bibinfo {author} {\bibfnamefont {G.}~\bibnamefont
  {Rajagopal}},\ }\href {\doibase 10.1103/PhysRevB.59.1917} {\bibfield
  {journal} {\bibinfo  {journal} {Phys. Rev. B}\ }\textbf {\bibinfo {volume}
  {59}},\ \bibinfo {pages} {1917} (\bibinfo {year} {1999})}\BibitemShut
  {NoStop}%
\bibitem [{\citenamefont {Drummond}\ \emph {et~al.}(2008)\citenamefont
  {Drummond}, \citenamefont {Needs}, \citenamefont {Sorouri},\ and\
  \citenamefont {Foulkes}}]{drummond_finite-size_2008}%
  \BibitemOpen
  \bibfield  {author} {\bibinfo {author} {\bibfnamefont {N.~D.}\ \bibnamefont
  {Drummond}}, \bibinfo {author} {\bibfnamefont {R.~J.}\ \bibnamefont {Needs}},
  \bibinfo {author} {\bibfnamefont {A.}~\bibnamefont {Sorouri}}, \ and\
  \bibinfo {author} {\bibfnamefont {W.~M.~C.}\ \bibnamefont {Foulkes}},\ }\href
  {\doibase 10.1103/PhysRevB.78.125106} {\bibfield  {journal} {\bibinfo
  {journal} {Phys. Rev. B}\ }\textbf {\bibinfo {volume} {78}},\ \bibinfo
  {pages} {125106} (\bibinfo {year} {2008})}\BibitemShut {NoStop}%
\bibitem [{\citenamefont {Chiesa}\ \emph {et~al.}(2006)\citenamefont {Chiesa},
  \citenamefont {Ceperley}, \citenamefont {Martin},\ and\ \citenamefont
  {Holzmann}}]{chiesa_finite-size_2006}%
  \BibitemOpen
  \bibfield  {author} {\bibinfo {author} {\bibfnamefont {S.}~\bibnamefont
  {Chiesa}}, \bibinfo {author} {\bibfnamefont {D.~M.}\ \bibnamefont
  {Ceperley}}, \bibinfo {author} {\bibfnamefont {R.~M.}\ \bibnamefont
  {Martin}}, \ and\ \bibinfo {author} {\bibfnamefont {M.}~\bibnamefont
  {Holzmann}},\ }\href {\doibase 10.1103/PhysRevLett.97.076404} {\bibfield
  {journal} {\bibinfo  {journal} {Phys. Rev. Lett.}\ }\textbf {\bibinfo
  {volume} {97}},\ \bibinfo {pages} {076404} (\bibinfo {year}
  {2006})}\BibitemShut {NoStop}%
\bibitem [{\citenamefont {Lin}\ \emph {et~al.}(2001)\citenamefont {Lin},
  \citenamefont {Zong},\ and\ \citenamefont
  {Ceperley}}]{lin_twist-averaged_2001}%
  \BibitemOpen
  \bibfield  {author} {\bibinfo {author} {\bibfnamefont {C.}~\bibnamefont
  {Lin}}, \bibinfo {author} {\bibfnamefont {F.~H.}\ \bibnamefont {Zong}}, \
  and\ \bibinfo {author} {\bibfnamefont {D.~M.}\ \bibnamefont {Ceperley}},\
  }\href {\doibase 10.1103/PhysRevE.64.016702} {\bibfield  {journal} {\bibinfo
  {journal} {Phys. Rev. E}\ }\textbf {\bibinfo {volume} {64}},\ \bibinfo
  {pages} {016702} (\bibinfo {year} {2001})}\BibitemShut {NoStop}%
\bibitem [{\citenamefont {Krogel}(2016)}]{krogel_nexus_2016}%
  \BibitemOpen
  \bibfield  {author} {\bibinfo {author} {\bibfnamefont {J.~T.}\ \bibnamefont
  {Krogel}},\ }\href {\doibase 10.1016/j.cpc.2015.08.012} {\bibfield  {journal}
  {\bibinfo  {journal} {Comput. Phys. Commun.}\ }\textbf {\bibinfo {volume}
  {198}},\ \bibinfo {pages} {154} (\bibinfo {year} {2016})}\BibitemShut
  {NoStop}%
\bibitem [{\citenamefont {Blum}\ \emph {et~al.}(2009)\citenamefont {Blum},
  \citenamefont {Gehrke}, \citenamefont {Hanke}, \citenamefont {Havu},
  \citenamefont {Havu}, \citenamefont {Ren}, \citenamefont {Reuter},\ and\
  \citenamefont {Scheffler}}]{blum2009ab}%
  \BibitemOpen
  \bibfield  {author} {\bibinfo {author} {\bibfnamefont {V.}~\bibnamefont
  {Blum}}, \bibinfo {author} {\bibfnamefont {R.}~\bibnamefont {Gehrke}},
  \bibinfo {author} {\bibfnamefont {F.}~\bibnamefont {Hanke}}, \bibinfo
  {author} {\bibfnamefont {P.}~\bibnamefont {Havu}}, \bibinfo {author}
  {\bibfnamefont {V.}~\bibnamefont {Havu}}, \bibinfo {author} {\bibfnamefont
  {X.}~\bibnamefont {Ren}}, \bibinfo {author} {\bibfnamefont {K.}~\bibnamefont
  {Reuter}}, \ and\ \bibinfo {author} {\bibfnamefont {M.}~\bibnamefont
  {Scheffler}},\ }\href@noop {} {\bibfield  {journal} {\bibinfo  {journal}
  {Computer Physics Communications}\ }\textbf {\bibinfo {volume} {180}},\
  \bibinfo {pages} {2175} (\bibinfo {year} {2009})}\BibitemShut {NoStop}%
\bibitem [{\citenamefont {Ren}\ \emph {et~al.}(2012)\citenamefont {Ren},
  \citenamefont {Rinke}, \citenamefont {Blum}, \citenamefont {Wieferink},
  \citenamefont {Tkatchenko}, \citenamefont {Sanfilippo}, \citenamefont
  {Reuter},\ and\ \citenamefont {Scheffler}}]{ren2012}%
  \BibitemOpen
  \bibfield  {author} {\bibinfo {author} {\bibfnamefont {X.}~\bibnamefont
  {Ren}}, \bibinfo {author} {\bibfnamefont {P.}~\bibnamefont {Rinke}}, \bibinfo
  {author} {\bibfnamefont {V.}~\bibnamefont {Blum}}, \bibinfo {author}
  {\bibfnamefont {J.}~\bibnamefont {Wieferink}}, \bibinfo {author}
  {\bibfnamefont {A.}~\bibnamefont {Tkatchenko}}, \bibinfo {author}
  {\bibfnamefont {A.}~\bibnamefont {Sanfilippo}}, \bibinfo {author}
  {\bibfnamefont {K.}~\bibnamefont {Reuter}}, \ and\ \bibinfo {author}
  {\bibfnamefont {M.}~\bibnamefont {Scheffler}},\ }\href@noop {} {\bibfield
  {journal} {\bibinfo  {journal} {New Journal of Physics}\ }\textbf {\bibinfo
  {volume} {14}},\ \bibinfo {pages} {053020} (\bibinfo {year}
  {2012})}\BibitemShut {NoStop}%
\bibitem [{\citenamefont {Marek}\ \emph {et~al.}(2014)\citenamefont {Marek},
  \citenamefont {Blum}, \citenamefont {Johanni}, \citenamefont {Havu},
  \citenamefont {Lang}, \citenamefont {Auckenthaler}, \citenamefont {Heinecke},
  \citenamefont {Bungartz},\ and\ \citenamefont {Lederer}}]{marek2014elpa}%
  \BibitemOpen
  \bibfield  {author} {\bibinfo {author} {\bibfnamefont {A.}~\bibnamefont
  {Marek}}, \bibinfo {author} {\bibfnamefont {V.}~\bibnamefont {Blum}},
  \bibinfo {author} {\bibfnamefont {R.}~\bibnamefont {Johanni}}, \bibinfo
  {author} {\bibfnamefont {V.}~\bibnamefont {Havu}}, \bibinfo {author}
  {\bibfnamefont {B.}~\bibnamefont {Lang}}, \bibinfo {author} {\bibfnamefont
  {T.}~\bibnamefont {Auckenthaler}}, \bibinfo {author} {\bibfnamefont
  {A.}~\bibnamefont {Heinecke}}, \bibinfo {author} {\bibfnamefont {H.-J.}\
  \bibnamefont {Bungartz}}, \ and\ \bibinfo {author} {\bibfnamefont
  {H.}~\bibnamefont {Lederer}},\ }\href@noop {} {\bibfield  {journal} {\bibinfo
   {journal} {Journal of Physics: Condensed Matter}\ }\textbf {\bibinfo
  {volume} {26}},\ \bibinfo {pages} {213201} (\bibinfo {year}
  {2014})}\BibitemShut {NoStop}%
\bibitem [{\citenamefont {Adamo}\ and\ \citenamefont {Barone}(1999)}]{PBE0}%
  \BibitemOpen
  \bibfield  {author} {\bibinfo {author} {\bibfnamefont {C.}~\bibnamefont
  {Adamo}}\ and\ \bibinfo {author} {\bibfnamefont {V.}~\bibnamefont {Barone}},\
  }\href {\doibase 10.1063/1.478522} {\bibfield  {journal} {\bibinfo  {journal}
  {The Journal of Chemical Physics}\ }\textbf {\bibinfo {volume} {110}},\
  \bibinfo {pages} {6158} (\bibinfo {year} {1999})},\ \Eprint
  {http://arxiv.org/abs/https://doi.org/10.1063/1.478522}
  {https://doi.org/10.1063/1.478522} \BibitemShut {NoStop}%
\bibitem [{DIR()}]{DIRAC19}%
  \BibitemOpen
  \href@noop {} {}\bibinfo {note} {{DIRAC}, a relativistic ab initio electronic
  structure program, Release {DIRAC19} (2019), written by A.~S.~P.~Gomes,
  T.~Saue, L.~Visscher, H.~J.~{\relax Aa}.~Jensen, and R.~Bast, with
  contributions from I.~A.~Aucar, V.~Bakken, K.~G.~Dyall, S.~Dubillard,
  U.~Ekstr{\"o}m, E.~Eliav, T.~Enevoldsen, E.~Fa{\ss}hauer, T.~Fleig,
  O.~Fossgaard, L.~Halbert, E.~D.~Hedeg{\aa}rd, B.~Heimlich--Paris,
  T.~Helgaker, J.~Henriksson, M.~Ilia{\v{s}}, Ch.~R.~Jacob, S.~Knecht,
  S.~Komorovsk{\'y}, O.~Kullie, J.~K.~L{\ae}rdahl, C.~V.~Larsen, Y.~S.~Lee,
  H.~S.~Nataraj, M.~K.~Nayak, P.~Norman, G.~Olejniczak, J.~Olsen,
  J.~M.~H.~Olsen, Y.~C.~Park, J.~K.~Pedersen, M.~Pernpointner, R.~di~Remigio,
  K.~Ruud, P.~Sa{\l}ek, B.~Schimmelpfennig, B.~Senjean, A.~Shee, J.~Sikkema,
  A.~J.~Thorvaldsen, J.~Thyssen, J.~van~Stralen, M.~L.~Vidal, S.~Villaume,
  O.~Visser, T.~Winther, and S.~Yamamoto (available at
  \url{http://dx.doi.org/10.5281/zenodo.3572669}, see also
  \url{http://www.diracprogram.org})}\BibitemShut {NoStop}%
\bibitem [{\citenamefont {Saue}\ \emph {et~al.}(2020)\citenamefont {Saue},
  \citenamefont {Bast}, \citenamefont {Gomes}, \citenamefont {Jensen},
  \citenamefont {Visscher}, \citenamefont {Aucar}, \citenamefont {Di~Remigio},
  \citenamefont {Dyall}, \citenamefont {Eliav}, \citenamefont {Fasshauer},
  \citenamefont {Fleig}, \citenamefont {Halbert}, \citenamefont {Hedegård},
  \citenamefont {Helmich-Paris}, \citenamefont {Iliaš}, \citenamefont {Jacob},
  \citenamefont {Knecht}, \citenamefont {Laerdahl}, \citenamefont {Vidal},
  \citenamefont {Nayak}, \citenamefont {Olejniczak}, \citenamefont {Olsen},
  \citenamefont {Pernpointner}, \citenamefont {Senjean}, \citenamefont {Shee},
  \citenamefont {Sunaga},\ and\ \citenamefont {van Stralen}}]{saue_dirac_2020}%
  \BibitemOpen
  \bibfield  {author} {\bibinfo {author} {\bibfnamefont {T.}~\bibnamefont
  {Saue}}, \bibinfo {author} {\bibfnamefont {R.}~\bibnamefont {Bast}}, \bibinfo
  {author} {\bibfnamefont {A.~S.~P.}\ \bibnamefont {Gomes}}, \bibinfo {author}
  {\bibfnamefont {H.~J.~A.}\ \bibnamefont {Jensen}}, \bibinfo {author}
  {\bibfnamefont {L.}~\bibnamefont {Visscher}}, \bibinfo {author}
  {\bibfnamefont {I.~A.}\ \bibnamefont {Aucar}}, \bibinfo {author}
  {\bibfnamefont {R.}~\bibnamefont {Di~Remigio}}, \bibinfo {author}
  {\bibfnamefont {K.~G.}\ \bibnamefont {Dyall}}, \bibinfo {author}
  {\bibfnamefont {E.}~\bibnamefont {Eliav}}, \bibinfo {author} {\bibfnamefont
  {E.}~\bibnamefont {Fasshauer}}, \bibinfo {author} {\bibfnamefont
  {T.}~\bibnamefont {Fleig}}, \bibinfo {author} {\bibfnamefont
  {L.}~\bibnamefont {Halbert}}, \bibinfo {author} {\bibfnamefont {E.~D.}\
  \bibnamefont {Hedegård}}, \bibinfo {author} {\bibfnamefont {B.}~\bibnamefont
  {Helmich-Paris}}, \bibinfo {author} {\bibfnamefont {M.}~\bibnamefont
  {Iliaš}}, \bibinfo {author} {\bibfnamefont {C.~R.}\ \bibnamefont {Jacob}},
  \bibinfo {author} {\bibfnamefont {S.}~\bibnamefont {Knecht}}, \bibinfo
  {author} {\bibfnamefont {J.~K.}\ \bibnamefont {Laerdahl}}, \bibinfo {author}
  {\bibfnamefont {M.~L.}\ \bibnamefont {Vidal}}, \bibinfo {author}
  {\bibfnamefont {M.~K.}\ \bibnamefont {Nayak}}, \bibinfo {author}
  {\bibfnamefont {M.}~\bibnamefont {Olejniczak}}, \bibinfo {author}
  {\bibfnamefont {J.~M.~H.}\ \bibnamefont {Olsen}}, \bibinfo {author}
  {\bibfnamefont {M.}~\bibnamefont {Pernpointner}}, \bibinfo {author}
  {\bibfnamefont {B.}~\bibnamefont {Senjean}}, \bibinfo {author} {\bibfnamefont
  {A.}~\bibnamefont {Shee}}, \bibinfo {author} {\bibfnamefont {A.}~\bibnamefont
  {Sunaga}}, \ and\ \bibinfo {author} {\bibfnamefont {J.~N.~P.}\ \bibnamefont
  {van Stralen}},\ }\href {\doibase 10.1063/5.0004844} {\bibfield  {journal}
  {\bibinfo  {journal} {J. Chem. Phys.}\ }\textbf {\bibinfo {volume} {152}},\
  \bibinfo {pages} {204104} (\bibinfo {year} {2020})},\ \bibinfo {note}
  {publisher: American Institute of Physics}\BibitemShut {NoStop}%
\bibitem [{\citenamefont {Chang}\ and\ \citenamefont
  {Wagner}(2020)}]{chang_effective_2020}%
  \BibitemOpen
  \bibfield  {author} {\bibinfo {author} {\bibfnamefont {Y.}~\bibnamefont
  {Chang}}\ and\ \bibinfo {author} {\bibfnamefont {L.~K.}\ \bibnamefont
  {Wagner}},\ }\href {\doibase 10.1103/PhysRevResearch.2.013195} {\bibfield
  {journal} {\bibinfo  {journal} {Physical Review Research}\ }\textbf {\bibinfo
  {volume} {2}},\ \bibinfo {pages} {013195} (\bibinfo {year} {2020})},\
  \bibinfo {note} {publisher: American Physical Society}\BibitemShut {NoStop}%
\bibitem [{\citenamefont {Annaberdiyev}\ \emph {et~al.}(2022)\citenamefont
  {Annaberdiyev}, \citenamefont {Melton}, \citenamefont {Wang},\ and\
  \citenamefont {Mitas}}]{annaberdiyev_electronic_2022}%
  \BibitemOpen
  \bibfield  {author} {\bibinfo {author} {\bibfnamefont {A.}~\bibnamefont
  {Annaberdiyev}}, \bibinfo {author} {\bibfnamefont {C.~A.}\ \bibnamefont
  {Melton}}, \bibinfo {author} {\bibfnamefont {G.}~\bibnamefont {Wang}}, \ and\
  \bibinfo {author} {\bibfnamefont {L.}~\bibnamefont {Mitas}},\ }\href
  {http://arxiv.org/abs/2203.15949} {\bibfield  {journal} {\bibinfo  {journal}
  {arXiv:2203.15949 [cond-mat, physics:physics]}\ } (\bibinfo {year} {2022})},\
  \bibinfo {note} {arXiv: 2203.15949}\BibitemShut {NoStop}%
\bibitem [{\citenamefont {Zen}\ \emph {et~al.}(2018)\citenamefont {Zen},
  \citenamefont {Brandenburg}, \citenamefont {Klimeš}, \citenamefont
  {Tkatchenko}, \citenamefont {Alfè},\ and\ \citenamefont
  {Michaelides}}]{zen_fast_2018}%
  \BibitemOpen
  \bibfield  {author} {\bibinfo {author} {\bibfnamefont {A.}~\bibnamefont
  {Zen}}, \bibinfo {author} {\bibfnamefont {J.~G.}\ \bibnamefont
  {Brandenburg}}, \bibinfo {author} {\bibfnamefont {J.}~\bibnamefont
  {Klimeš}}, \bibinfo {author} {\bibfnamefont {A.}~\bibnamefont {Tkatchenko}},
  \bibinfo {author} {\bibfnamefont {D.}~\bibnamefont {Alfè}}, \ and\ \bibinfo
  {author} {\bibfnamefont {A.}~\bibnamefont {Michaelides}},\ }\href {\doibase
  10.1073/pnas.1715434115} {\bibfield  {journal} {\bibinfo  {journal} {Proc.
  Natl. Acad. Sci.}\ }\textbf {\bibinfo {volume} {115}},\ \bibinfo {pages}
  {1724} (\bibinfo {year} {2018})}\BibitemShut {NoStop}%
\bibitem [{\citenamefont {Melton}\ and\ \citenamefont
  {Mitas}(2020)}]{melton2020}%
  \BibitemOpen
  \bibfield  {author} {\bibinfo {author} {\bibfnamefont {C.}~\bibnamefont
  {Melton}}\ and\ \bibinfo {author} {\bibfnamefont {L.}~\bibnamefont {Mitas}},\
  }\href@noop {} {\bibfield  {journal} {\bibinfo  {journal} {Phys. Rev. B}\
  }\textbf {\bibinfo {volume} {102}},\ \bibinfo {pages} {045103} (\bibinfo
  {year} {2020})}\BibitemShut {NoStop}%
\bibitem [{\citenamefont {Annaberdiyev}\ \emph {et~al.}(2020)\citenamefont
  {Annaberdiyev}, \citenamefont {Wang}, \citenamefont {Melton}, \citenamefont
  {Bennett}, \citenamefont {Schulenburger},\ and\ \citenamefont
  {Mitas}}]{gani2020}%
  \BibitemOpen
  \bibfield  {author} {\bibinfo {author} {\bibfnamefont {A.}~\bibnamefont
  {Annaberdiyev}}, \bibinfo {author} {\bibfnamefont {G.}~\bibnamefont {Wang}},
  \bibinfo {author} {\bibfnamefont {C.}~\bibnamefont {Melton}}, \bibinfo
  {author} {\bibfnamefont {M.}~\bibnamefont {Bennett}}, \bibinfo {author}
  {\bibfnamefont {L.}~\bibnamefont {Schulenburger}}, \ and\ \bibinfo {author}
  {\bibfnamefont {L.}~\bibnamefont {Mitas}},\ }\href@noop {} {\bibfield
  {journal} {\bibinfo  {journal} {J. Chem. Theory Comp.}\ }\textbf {\bibinfo
  {volume} {16}},\ \bibinfo {pages} {1482} (\bibinfo {year}
  {2020})}\BibitemShut {NoStop}%
\bibitem [{\citenamefont {Nepal}\ \emph {et~al.}(2020)\citenamefont {Nepal},
  \citenamefont {Adhikari}, \citenamefont {Neupane},\ and\ \citenamefont
  {Ruzsinszky}}]{nepal_formation_2020}%
  \BibitemOpen
  \bibfield  {author} {\bibinfo {author} {\bibfnamefont {N.~K.}\ \bibnamefont
  {Nepal}}, \bibinfo {author} {\bibfnamefont {S.}~\bibnamefont {Adhikari}},
  \bibinfo {author} {\bibfnamefont {B.}~\bibnamefont {Neupane}}, \ and\
  \bibinfo {author} {\bibfnamefont {A.}~\bibnamefont {Ruzsinszky}},\ }\href
  {\doibase 10.1103/PhysRevB.102.205121} {\bibfield  {journal} {\bibinfo
  {journal} {Phys. Rev. B}\ }\textbf {\bibinfo {volume} {102}},\ \bibinfo
  {pages} {205121} (\bibinfo {year} {2020})}\BibitemShut {NoStop}%
\bibitem [{\citenamefont {Isaacs}\ \emph {et~al.}(2020)\citenamefont {Isaacs},
  \citenamefont {Patel},\ and\ \citenamefont
  {Wolverton}}]{isaacs_prediction_2020}%
  \BibitemOpen
  \bibfield  {author} {\bibinfo {author} {\bibfnamefont {E.~B.}\ \bibnamefont
  {Isaacs}}, \bibinfo {author} {\bibfnamefont {S.}~\bibnamefont {Patel}}, \
  and\ \bibinfo {author} {\bibfnamefont {C.}~\bibnamefont {Wolverton}},\ }\href
  {\doibase 10.1103/PhysRevMaterials.4.065405} {\bibfield  {journal} {\bibinfo
  {journal} {Phys. Rev. Materials}\ }\textbf {\bibinfo {volume} {4}},\ \bibinfo
  {pages} {065405} (\bibinfo {year} {2020})}\BibitemShut {NoStop}%
\bibitem [{\citenamefont {Blaiszik}\ \emph {et~al.}(2016)\citenamefont
  {Blaiszik}, \citenamefont {Chard}, \citenamefont {Pruyne}, \citenamefont
  {Ananthakrishnan}, \citenamefont {Tuecke},\ and\ \citenamefont
  {Foster}}]{blaiszik_materials_2016}%
  \BibitemOpen
  \bibfield  {author} {\bibinfo {author} {\bibfnamefont {B.}~\bibnamefont
  {Blaiszik}}, \bibinfo {author} {\bibfnamefont {K.}~\bibnamefont {Chard}},
  \bibinfo {author} {\bibfnamefont {J.}~\bibnamefont {Pruyne}}, \bibinfo
  {author} {\bibfnamefont {R.}~\bibnamefont {Ananthakrishnan}}, \bibinfo
  {author} {\bibfnamefont {S.}~\bibnamefont {Tuecke}}, \ and\ \bibinfo {author}
  {\bibfnamefont {I.}~\bibnamefont {Foster}},\ }\href {\doibase
  10.1007/s11837-016-2001-3} {\bibfield  {journal} {\bibinfo  {journal} {JOM}\
  }\textbf {\bibinfo {volume} {68}},\ \bibinfo {pages} {2045} (\bibinfo {year}
  {2016})}\BibitemShut {NoStop}%
\bibitem [{\citenamefont {Blaiszik}\ \emph {et~al.}(2019)\citenamefont
  {Blaiszik}, \citenamefont {Ward}, \citenamefont {Schwarting}, \citenamefont
  {Gaff}, \citenamefont {Chard}, \citenamefont {Pike}, \citenamefont {Chard},\
  and\ \citenamefont {Foster}}]{blaiszik_data_2019}%
  \BibitemOpen
  \bibfield  {author} {\bibinfo {author} {\bibfnamefont {B.}~\bibnamefont
  {Blaiszik}}, \bibinfo {author} {\bibfnamefont {L.}~\bibnamefont {Ward}},
  \bibinfo {author} {\bibfnamefont {M.}~\bibnamefont {Schwarting}}, \bibinfo
  {author} {\bibfnamefont {J.}~\bibnamefont {Gaff}}, \bibinfo {author}
  {\bibfnamefont {R.}~\bibnamefont {Chard}}, \bibinfo {author} {\bibfnamefont
  {D.}~\bibnamefont {Pike}}, \bibinfo {author} {\bibfnamefont {K.}~\bibnamefont
  {Chard}}, \ and\ \bibinfo {author} {\bibfnamefont {I.}~\bibnamefont
  {Foster}},\ }\href {\doibase 10.1557/mrc.2019.118} {\bibfield  {journal}
  {\bibinfo  {journal} {MRC}\ }\textbf {\bibinfo {volume} {9}},\ \bibinfo
  {pages} {1125} (\bibinfo {year} {2019})},\ \bibinfo {note} {arXiv:
  1904.10423}\BibitemShut {NoStop}%
\bibitem [{\citenamefont {Isaacs}\ \emph {et~al.}(2022)\citenamefont {Isaacs},
  \citenamefont {Shin}, \citenamefont {Annaberdiyev}, \citenamefont
  {Wolverton}, \citenamefont {Mitas}, \citenamefont {Benali},\ and\
  \citenamefont {Heinonen}}]{mdf_data}%
  \BibitemOpen
  \bibfield  {author} {\bibinfo {author} {\bibfnamefont {E.~B.}\ \bibnamefont
  {Isaacs}}, \bibinfo {author} {\bibfnamefont {H.}~\bibnamefont {Shin}},
  \bibinfo {author} {\bibfnamefont {A.}~\bibnamefont {Annaberdiyev}}, \bibinfo
  {author} {\bibfnamefont {C.}~\bibnamefont {Wolverton}}, \bibinfo {author}
  {\bibfnamefont {L.}~\bibnamefont {Mitas}}, \bibinfo {author} {\bibfnamefont
  {A.}~\bibnamefont {Benali}}, \ and\ \bibinfo {author} {\bibfnamefont
  {O.}~\bibnamefont {Heinonen}},\ }\href {\doibase 10.18126/359I-7W2P}
  {\bibfield  {journal} {\bibinfo  {journal} {Materials Data Facility}\ }
  (\bibinfo {year} {2022}),\ 10.18126/359I-7W2P}\BibitemShut {NoStop}%
\bibitem [{\citenamefont {Makov}\ and\ \citenamefont
  {Payne}(1995)}]{makov_periodic_1995}%
  \BibitemOpen
  \bibfield  {author} {\bibinfo {author} {\bibfnamefont {G.}~\bibnamefont
  {Makov}}\ and\ \bibinfo {author} {\bibfnamefont {M.~C.}\ \bibnamefont
  {Payne}},\ }\href {\doibase 10.1103/PhysRevB.51.4014} {\bibfield  {journal}
  {\bibinfo  {journal} {Phys. Rev. B}\ }\textbf {\bibinfo {volume} {51}},\
  \bibinfo {pages} {4014} (\bibinfo {year} {1995})}\BibitemShut {NoStop}%
\bibitem [{\citenamefont {Kramida}\ \emph {et~al.}(2021)\citenamefont
  {Kramida}, \citenamefont {{Yu.~Ralchenko}}, \citenamefont {Reader},\ and\
  \citenamefont {{and NIST ASD Team}}}]{NIST_ASD}%
  \BibitemOpen
  \bibfield  {author} {\bibinfo {author} {\bibfnamefont {A.}~\bibnamefont
  {Kramida}}, \bibinfo {author} {\bibnamefont {{Yu.~Ralchenko}}}, \bibinfo
  {author} {\bibfnamefont {J.}~\bibnamefont {Reader}}, \ and\ \bibinfo {author}
  {\bibnamefont {{and NIST ASD Team}}},\ }\href@noop {} {}\bibinfo
  {howpublished} {{NIST Atomic Spectra Database (ver. 5.9), [Online].
  Available: {\tt{https://physics.nist.gov/asd}} [2022, April 1]. National
  Institute of Standards and Technology, Gaithersburg, MD.}} (\bibinfo {year}
  {2021})\BibitemShut {NoStop}%
\bibitem [{\citenamefont {Al-Saidi}(2008)}]{al-saidi_variational_2008}%
  \BibitemOpen
  \bibfield  {author} {\bibinfo {author} {\bibfnamefont {W.~A.}\ \bibnamefont
  {Al-Saidi}},\ }\href {\doibase 10.1063/1.2969098} {\bibfield  {journal}
  {\bibinfo  {journal} {J. Chem. Phys.}\ }\textbf {\bibinfo {volume} {129}},\
  \bibinfo {pages} {064316} (\bibinfo {year} {2008})}\BibitemShut {NoStop}%
\bibitem [{\citenamefont {Peterson}\ \emph {et~al.}(2003)\citenamefont
  {Peterson}, \citenamefont {Figgen}, \citenamefont {Goll}, \citenamefont
  {Stoll},\ and\ \citenamefont {Dolg}}]{peterson_systematically_2003}%
  \BibitemOpen
  \bibfield  {author} {\bibinfo {author} {\bibfnamefont {K.~A.}\ \bibnamefont
  {Peterson}}, \bibinfo {author} {\bibfnamefont {D.}~\bibnamefont {Figgen}},
  \bibinfo {author} {\bibfnamefont {E.}~\bibnamefont {Goll}}, \bibinfo {author}
  {\bibfnamefont {H.}~\bibnamefont {Stoll}}, \ and\ \bibinfo {author}
  {\bibfnamefont {M.}~\bibnamefont {Dolg}},\ }\href {\doibase
  10.1063/1.1622924} {\bibfield  {journal} {\bibinfo  {journal} {J. Chem.
  Phys.}\ }\textbf {\bibinfo {volume} {119}},\ \bibinfo {pages} {11113}
  (\bibinfo {year} {2003})}\BibitemShut {NoStop}%
\bibitem [{\citenamefont {Wang}\ \emph {et~al.}(2022)\citenamefont {Wang},
  \citenamefont {Kincaid}, \citenamefont {Zhou}, \citenamefont {Annaberdiyev},
  \citenamefont {Bennett}, \citenamefont {Krogel},\ and\ \citenamefont
  {Mitas}}]{wang_new_2022}%
  \BibitemOpen
  \bibfield  {author} {\bibinfo {author} {\bibfnamefont {G.}~\bibnamefont
  {Wang}}, \bibinfo {author} {\bibfnamefont {B.}~\bibnamefont {Kincaid}},
  \bibinfo {author} {\bibfnamefont {H.}~\bibnamefont {Zhou}}, \bibinfo {author}
  {\bibfnamefont {A.}~\bibnamefont {Annaberdiyev}}, \bibinfo {author}
  {\bibfnamefont {M.~C.}\ \bibnamefont {Bennett}}, \bibinfo {author}
  {\bibfnamefont {J.~T.}\ \bibnamefont {Krogel}}, \ and\ \bibinfo {author}
  {\bibfnamefont {L.}~\bibnamefont {Mitas}},\ }\href
  {http://arxiv.org/abs/2202.04747} {\bibfield  {journal} {\bibinfo  {journal}
  {arXiv:2202.04747 [cond-mat, physics:physics]}\ } (\bibinfo {year} {2022})},\
  \bibinfo {note} {arXiv: 2202.04747}\BibitemShut {NoStop}%
\bibitem [{\citenamefont {Yang}\ \emph {et~al.}(2016)\citenamefont {Yang},
  \citenamefont {Kneiß}, \citenamefont {Schein}, \citenamefont {Lorenz},\ and\
  \citenamefont {Grundmann}}]{yang_room-temperature_2016}%
  \BibitemOpen
  \bibfield  {author} {\bibinfo {author} {\bibfnamefont {C.}~\bibnamefont
  {Yang}}, \bibinfo {author} {\bibfnamefont {M.}~\bibnamefont {Kneiß}},
  \bibinfo {author} {\bibfnamefont {F.-L.}\ \bibnamefont {Schein}}, \bibinfo
  {author} {\bibfnamefont {M.}~\bibnamefont {Lorenz}}, \ and\ \bibinfo {author}
  {\bibfnamefont {M.}~\bibnamefont {Grundmann}},\ }\href {\doibase
  10.1038/srep21937} {\bibfield  {journal} {\bibinfo  {journal} {Scientific
  Reports}\ }\textbf {\bibinfo {volume} {6}},\ \bibinfo {pages} {21937}
  (\bibinfo {year} {2016})},\ \bibinfo {note} {number: 1 Publisher: Nature
  Publishing Group}\BibitemShut {NoStop}%
\end{thebibliography}%

\clearpage
\appendix

\renewcommand{\arraystretch}{1.2}

\section{Tabulated Formation Energies}

Tables \ref{tab:CuI} and \ref{tab:VPt2} provide the formation energies presented in the main portion of the paper.

\begin{table}[!htbp]
\centering
\caption{
Formation energies for CuI in 
units of meV/atom.
}
\label{tab:CuI}
\begin{tabular}{l|ccccc}
\hline
Method  & Energy \\
\hline
PBE     & -156 \\
SCAN    & -151 \\
PBE0    & -419 \\
PBE0+SO & -401 \\
DMC     & -511(50) \\
DMC+SO  & -493(50) \\
Expt.   & -369(20) \\
\hline
\end{tabular}
\end{table}

\begin{table}[!htbp]
\centering
\caption{
Formation energies for VPt$_2$ in 
units of meV/atom.
}
\label{tab:VPt2}
\begin{tabular}{l|ccccc}
\hline
Method  & Energy \\
\hline
PBE     & -555 \\
SCAN    & -737 \\
PBE0    & -759\\
PBE0+SO & -720\\
DMC     & -764(50)\\
DMC+SO  & -725(50)\\
Expt.   & -386\\
\hline
\end{tabular}
\end{table}

\section{Validation of Iodine pseudopotential}

\subsection{Scalar Relativistic Terms}
Here, we discuss atomic calculations to validate the
Burkatzki-Filippi-Dolg (BFD) iodine pseudopotential \cite{burkatzki_energy-consistent_2007}. We use a cubic
simulation cell with 20~\AA\ side length and employ the Makov-Payne
finite-size correction \cite{makov_periodic_1995}. 
The first two ionization potentials (IPs) and the first electron affinity (EA) are tested.
The atomic configurations, taken from the NIST Atomic Spectra
Database \cite{NIST_ASD}, are $^1$S$_0$ for the $-1$ charge state, $^2$P$^o_{3/2}$ for
the $0$ state, $^3$P$_2$ for the $+1$ state, and $^4$S$_{1/2}$ for the
$+2$ state.

\begin{table*}[htbp]
  \begin{tabular}{c|c|c|c|c|c|c}
    & This work DFT & This work DMC & Past DFT\cite{al-saidi_variational_2008} & Past DMC & Experiment & AE CCSD(T) \\ 
    \hline
      IP1 & 10.49 & 10.51 & 10.51 & 10.57              & 10.45 & 10.5697 \\
      IP2 & 18.69 & 19.95 &   --  & --                 & 19.13 & 18.7665 \\
      EA  &  3.24 &  3.33 & 3.25  & 3.34               &  3.06 &  3.2923 \\
  \end{tabular}
    \begin{centering}
      \caption{Comparison of atomic iodine ionization potential and
        electronic affinity (in units of eV) results.
        AE CCSD(T) represents scalar relativistic, fully correlated, all-electron CCSD(T) calculation results.
        }
    \end{centering}
    \label{tab:iodine_pseudo}
\end{table*}

As shown in Table~\ref{tab:iodine_pseudo}, the computed IPs and EA
agree well with past work, AE CCSD(T), and experiment, validating the
use of this pseudopotential. 
Note that the experimental value of 3.06~eV for EA must be corrected for spin-orbit interactions for a direct comparison.
The average of atomic multiplets provides a value of 3.37~eV (see Refs. \citenum{al-saidi_variational_2008} and \citenum{peterson_systematically_2003}).
A more thorough transferability tests of BFD iodine ECP can be found in Ref. \citenum{wang_new_2022}.

\subsection{Spin-Orbit Terms}

Because the BFD pseudopotential lacks the SO terms, we adopted the STU SO terms \cite{STU} and tested the transferability with the results summarized in Table~\ref{tab:I_sorep}.
We can see that almost the same quality as from STU pseudopotential is obtained, validating the direct use of STU SO terms.
These tests also suggest that the STU SO terms can be adopted without changes for other elements.
Analogously, STU SO terms were added to Pt pseudopotential for consistency.

\begin{table}[!htbp]
\setlength{\tabcolsep}{6pt} 
\small
\centering
\caption{
Iodine atomic excitation errors for STU versus BFD in SOREP forms.
The errors are shown for full-relativistic X2C AE gaps using COSCI.
All values are in eV.
STU SO terms were used for BFD calculations.
}
\label{tab:I_sorep}
\begin{tabular}{llr|rrrr}
\hline
State & Term & AE & STU & BFD \\
\hline
$5s^25p^5$ & $^{2}P_{3/2}$ &  0.000 &      0.000 &        0.000  \\ 
$5s^25p^6$ & $^{1}S_{0}$   & -2.185 &      0.022 &        0.025  \\ 
$5s^25p^4$ & $^{3}P_{2}$   &  9.413 &      0.009 &        0.006  \\ 
$5s^25p^3$ & $^{4}S_{3/2}$ & 27.262 &      0.097 &        0.092  \\ 
\hline
$5s^25p^5$ & $^{2}P_{3/2}$ &  0.000 &      0.000 &        0.000  \\ 
        {} & $^{2}P_{1/2}$ &  0.962 &      0.030 &        0.024  \\ 
\hline
$5s^25p^4$ & $^{3}P_{2}$   &  0.000 &      0.000 &        0.000  \\ 
        {} & $^{3}P_{0}$   &  0.863 &      0.046 &        0.041  \\ 
        {} & $^{3}P_{1}$   &  0.869 &      0.019 &        0.015  \\ 
        {} & $^{1}D_{2}$   &  1.999 &     -0.029 &       -0.036  \\ 
        {} & $^{1}S_{0}$   &  4.258 &     -0.104 &       -0.119  \\ 
\hline
$5s^25p^3$ & $^{4}S_{3/2}$ &  0.000 &      0.000 &        0.000  \\ 
        {} & $^{2}D_{3/2}$ &  1.959 &     -0.136 &       -0.142  \\ 
        {} & $^{2}D_{5/2}$ &  2.348 &     -0.101 &       -0.109  \\ 
        {} & $^{2}P_{1/2}$ &  3.765 &     -0.183 &       -0.195  \\ 
        {} & $^{2}P_{3/2}$ &  4.375 &     -0.116 &       -0.131  \\ 
\hline        
MAD        &               &        &      0.074 &        0.078  \\ 

\hline
\end{tabular}
\end{table}

\section{Finite size extrapolation for solid I$_2$}

\begin{figure*}[htbp]
  \begin{center}
    \includegraphics[width=0.80\linewidth]{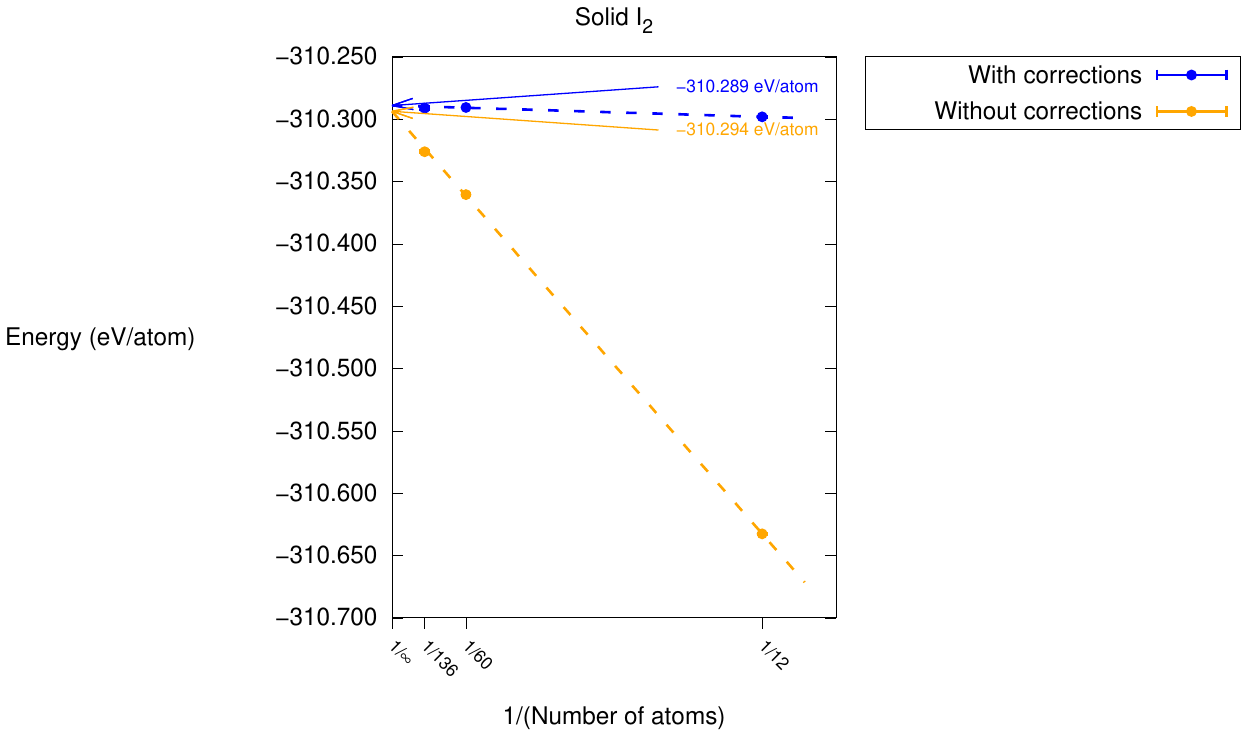}
  \end{center}
  \caption{Finite size extrapolation for solid I$_2$ from calculations
    with and without finite size corrections}
  \label{fig:iodine_extrap}
\end{figure*}

A comparison of the finite size extrapolation of solid I$_2$ with and
without finite size corrections (model periodic Coulomb interaction
and Chiesa-Ceperley-Martin-Holzmann kinetic energy correction) is
shown in Fig. \ref{fig:iodine_extrap}.
Both cases lead to very similar
values for the bulk limit, using linear extrapolation. We note that
the energy per atom does not depend strongly on the supercell when
the finite size corrections are included, as discussed in the main
text.

\section{Alternative estimation of SO effect in CuI}

In addition to the direct PBE0+SO effects used in the paper, here we use a different way to estimate the SO effect.
In order to evaluate the impact of spin-orbit interactions on CuI, we use
\begin{equation}
\Delta^{SO}=\Delta^{SO}({\rm CuI},s)-(1/2)\Delta^{SO}({\rm I}_2,s).  
\end{equation}
The first contribution is the shift for solid ($s$) CuI where we employ
 the value for the CuI molecule ($m$) to minimize the systematic
 biases. This is arranged into several differences that maximize the error cancellation as follows:
\begin{flalign}
&\Delta^{SO}({\rm CuI},s)  & {}\nonumber\\
& = \Delta^{SO}({\rm CuI},s)
\pm \Delta^{SO}({\rm CuI},m)\nonumber\\
 & =  [\Delta_{coh}^{SO}({\rm CuI},s)-\Delta_{bind}^{SO}({\rm CuI},m)]_{\rm DFT}+
\Delta^{SO}({\rm CuI},m)\nonumber\\
 &\approx  63 -37 -36 = -10\; {\rm meV/atom},
\end{flalign}
where we use the DFT/PBE0 values of cohesion and binding while the total energy shift $\Delta^{so}({\rm CuI},m)$ is found by COSCI calculations.  All values above are reasonably small, suggesting a very similar impact of spin-orbit in both ($s,m$) systems so that the use of DFT is justifiable. Note that the spin-orbit contributions to the DFT atomic energies, which are significant and can be biased due to difficulties in imposing correct atomic symmetries, cancel out in the difference (solid cohesion $-$ molecule binding).
On the other hand, the shift in the total energy of the molecule is evaluated more precisely, 
using the variational COSCI method.
Overall, the very small resulting value is easy to understand qualitatively since CuI solid is basically an ionic compound. 
The anion I$^-$ has a closed shell $p^6$, resulting in a very small net spin-orbit induced difference. This contrasts with the corresponding shift for the valence-only, neutral, isolated atom which is approximately
$-$362 meV.

For the  second contribution, we assume 
that the I$_2$ molecule and solid exhibit basically the same spin-orbit shifts and
we write:
$$
(1/2)\Delta^{SO}({\rm I}_2,s)\approx (1/2)\Delta^{SO}({\rm I}_2,m)= -47\; {\rm meV/atom}
$$
where we used value from the CCSD(T) calculation.
The full correction is then:
$$
\Delta^{SO}=\Delta^{SO}({\rm CuI},s)- (1/2)\Delta^{SO}({\rm I}_2,s)= 37\; {\rm meV/atom}
$$
which agrees reasonably with direct PBE0+SO calculation SO shift of 18 meV/atom.
The correction reduces the AREP formation energy of $-$511 meV/atom to 
$-$473 meV/atom which is reasonably close to the experimental value of  $-369\pm10$~meV/atom. The spin-orbit 
effect in the I$_2$ solid might be moderately larger as guessing from comparisons of binding ($m$) and cohesive ($s$) energies, reaching perhaps $\approx$ 50 meV/atom 
so that the remaining discrepancy might be possibly even smaller. In summary, correcting the QMC value based on calculations
incorporating spin-orbit interactions for CuI and solid I$_2$ brings theory and experiment into not perfect but a much improved agreement (within $\sim$ 100 meV/atom).

\section{QMC Variances}
Here we report the values for the
variance of the local energy, which would be identically zero for the
exact ground state. We find values normalized by the local energy to be similar, namely,
0.012 Ha for VPt$_2$, 0.023 Ha for bcc V, 0.009 Ha for fcc
Pt, 0.023 Ha for CuI, 0.023 Ha for fcc Cu, and 0.016 Ha for solid
I$_2$. The low variance values indicate very good consistency
of optimizations for Jastrow parameters across the calculated systems.

\section{CuI DFT Gaps}
Table \ref{tab:CuI_gap} provides the DFT and experimental gaps for CuI.
Note the excellent agreement of PBE0+SOC and the experimental value.

\begin{table}[!htbp]
\centering
\caption{
Band gaps for CuI in units of eV.
}
\label{tab:CuI_gap}
\begin{tabular}{l|ccccc}
\hline
Method  & Gap [eV] \\
\hline
PBE      & 1.25 \\
SCAN     & 3.64 \\
PBE0     & 3.38 \\
PBE0+SOC & 3.09 \\
Expt.\cite{yang_room-temperature_2016, grundmann_cuprous_2013} 
         & 3.1 \\
\hline
\end{tabular}
\end{table}
\end{document}